\documentclass[a4paper,fleqn]{cas-dc}

\usepackage{graphicx,subcaption}
\usepackage{rotating}
\usepackage{lscape}
\usepackage[justification=centering]{caption}
\usepackage{layout}
\usepackage{caption}
\usepackage[numbers]{natbib}
\usepackage[utf8]{inputenc}
\usepackage{aas_macros}

\def\tsc#1{\csdef{#1}{\textsc{\lowercase{#1}}\xspace}}
\tsc{WGM}
\tsc{QE}

\begin{document}
\UseRawInputEncoding
\let\WriteBookmarks\relax
\def\floatpagepagefraction{1}
\def\textpagefraction{.001}

\shorttitle{$\gamma$-ray outburst of 4C 31.03}    

\shortauthors{A.Thekkoth et al}  

\title [mode = title]{Gamma-ray variability and multi-wavelength insights into the unprecedented outburst from 4C\,31.03}  



%

\author[1]{Aminabi Thekkoth}[orcid=0000-0000-0000-0000]

\cormark[1]


\ead{thekkothaminabi@gmail.com}



\affiliation[1]{organization={University of Calicut},
            addressline={Thenjippalam}, 
            city={Malappuram},
            postcode={673635}, 
            state={Kerala},
            country={India}}

\author[1]{Baheeja C}[]

\ead{baheeja314@gmail.com}

\author[2,3]{S Sahayanathan}[]
\ead{sunder@barc.gov.in}

\author[1]{C D Ravikumar}[]
\ead{cdr@uoc.ac.in}

\affiliation[2]{organization={Bhaba Atomic Research Centre},
            city={Mumbai},
            postcode={400085}, 
            state={Maharashtra},
            country={India}}
\affiliation[3]{organization={Homi Bhabha National Institute},
            city={Mumbai},
            postcode={400094}, 
            state={Maharashtra},
            country={India}}

\cortext[cor1]{Corresponding author}


\nonumnote{}

\begin{abstract}
The blazar 4C\,31.03 recently underwent a major $\gamma$-ray outburst at the beginning of 2023 after a prolonged quiescent phase. \emph{Fermi}-LAT reported a daily average flux of $5\,\times10^{-6}$ phs cm$^{-2}$ s$^{-1}$, which is about 60 times its average value. We investigated this extraordinary outbreak through temporal and multi-wavelength analysis. From the statistical analysis of the $\gamma$-ray lightcurves using Bayesian blocks, we identified 3 epochs of prominent flares. The fastest flux decay during this  major outburst was observed within $5.5\pm 0.7$ hours. 
 The highest energy of $\gamma$-ray photons found from the source during the active phase is $\sim 82$\,GeV. Using the transparency of $\gamma$-rays against pair production and light crossing time argument, we could obtain the minimum jet Doppler factor as 17 corresponding to the flaring state.
 The broadband spectral energy distribution study performed using synchrotron, SSC and EC emission processes supports the external Compton scattering of IR photons as the likely mechanism for the $\gamma$-ray emission from the source. The results of this study suggest the scenario of the emission region in 4C\,31.03, being located beyond the Broad line region from the central blackhole. Long-term $\gamma$-ray flux distribution of 4C\,31.03 depicts a double log-normal variability, indicating that two distinct flux states are active in this energy band. 
The index distribution also reveals a two distinct variability patterns, suggesting that the $\gamma$-ray spectrum can be more precisely described by two photon indices.
\end{abstract}

\begin{keywords}
 galaxies:active\sep quasars:individual(4C 31.03) \sep radiation mechanisms:non-thermal \sep method:data analysis \sep gamma rays
\end{keywords}

\maketitle

\section{\label{sec:intro}Introduction}
\emph{Fermi}-LAT is a $\gamma$-ray space telescope, that observes the cosmos in the energy range 20 MeV to 1 TeV. With a wide field of view, LAT scans the entire sky every 3 hours.
Recently, in January 2023, \emph{Fermi}-LAT reported an enhancement in the $\gamma$-ray activity from the source 4C\,31.03,\,consistent with the coordinates RA=18.2227 and DEC=32.1399 \cite{2023ATel15841....1C}. 4C\,31.03 is recognized as a flat spectrum radio quasar (FSRQ) (z=0.603, \cite{1976ApJS...31..143W}) and is included in the \emph{Fermi} 4FGL point source catalog with identifier J0112.8+3208 \cite{2020ApJS..247...33A}. This source is also historically known in other names such as, B2\,0110+31, OC\,317, and NRAO 62. Further, it is considered as a possible candidate for the study of neutrino emission from Blazars (\cite{2022ApJ...939..123C}). 
This recent high $\gamma$-ray outburst marks the first instance of such an extraordinary activity since its discovery. The daily average gamma-ray flux reported ($\sim\,4\times 10^{-6}$ phs cm$^{-2}$ s$^{-1}$) is 60 times greater than the average value quoted in the 4FGL catalog. 
Following this, the IceCube Collaboration has conducted a search for possible muon neutrino events from the source and obtained an upper limit on the neutrino flux, with a 90\% confidence level  \cite{2023ATel15889....1H}. After this flaring event, the source reverted back to its quiescent flux level within a few days. The source again underwent a major transition to the phase of high activity during June 2023 with a maximum flux of $4.9 \times 10^{-6}$ phs cm$^{-2}$ s$^{-1}$ \cite{2023ATel16068....1C}. The variability aspects of this source has not been thoroughly studied before, despite continuous monitoring by \emph{Fermi}-LAT since 2009.

Under the unified model of radio-loud active galactic nuclei, blazars form the subclass with their relativistic jet pointed towards the earth \cite{1995PASP..107..803U}. The relativistic effects boost the intrinsic emission from blazars making them extremely luminous in the universe. Their high luminosity in $\gamma$-rays makes them the primary target of \emph{Fermi}-LAT. 
FSRQs are the luminous category of blazars with maximum power emitted at gamma-ray energies \cite{2002A&A...386..833G,1998MNRAS.301..451G, 1999ApJS..123...79H}. These objects exhibit strong continuum emission lines in the optical spectrum whereas, BL\,Lacs have featureless optical spectra \cite{1995PASP..107..803U, 2013ApJ...764..135S}. The spectral energy distribution (SED) of blazars is non-thermal spanning from radio to gamma-rays and features two broad peaks.The frequency at which the low-energy SED component peaks varies among the blazar classes, with FSRQ's having the least peak energy falling around infrared frequency \cite{2002A&A...386..833G,2008MNRAS.387.1669G}.

The theoretical models which are used to interpret the observed broadband SED of blazars, fall mainly under two emission scenarios namely leptonic and/or hadronic origin \cite{Cerruti_2020}. Under both these scenarios, the lower energy spectral component has been well understood to be synchrotron emission from a relativistic population of non-thermal electrons in the jet \cite{2012MNRAS.419.1660S, 1986rpa..book.....R}. However, the underlying mechanism for the higher-energy (HE) emission and the location of the main emission region from the central blackhole, still remain as open questions \cite{2013EPJWC..6105004J}.
Leptonic models interpret the higher energy SED via inverse Compton scattering of soft photons which can be synchrotron photon itself (SSC) and/or the photon fields external to the jet (EC) \cite{Sikora_94, Dermer_1994}. The ambient external photon field can be Ly-$\alpha$ line emission from broad line region (BLR) or the thermal IR emission from molecular torus. In the case of FSRQs, The X-ray emission is generally attributed to SSC, while the $\gamma$-ray emission is better explained by the EC process \cite{2012MNRAS.419.1660S, 2017MNRAS.470.3283S}. 
Identification of the target photon field for the EC process can also provide hints about the location of the emission region in FSRQs \cite{2013EPJWC..6105004J, 2009ApJ...692...32D}. Hadronic models, on the other hand, explain the higher energy SED component as either proton synchrotron or the emission due to nuclear cascades \cite{2000NewA....5..377A, 2011sf2a.conf..555C, 1990ApJ...362...38B}. 

Another intriguing feature of blazars is their variability patterns on timescales ranging from few minutes to days \cite{Abdo_2010, 2015ApJ...808L..48P, 2010MNRAS.405L..94T}. 
The availability of continuous observations from \emph{Fermi}-LAT presents a surfeit of data to search for the shortest time variability events.
The knowledge of the shortest variability time ($\tau$) has been often used to constrain the emission region size and its location in blazar jets. The light crossing time suggests the emission region size, $R$, is less than $c\,\delta\,\tau/(1+z)$, where, $\delta$ is the relativistic Doppler factor. If the emission arises from a homogeneous region, then $R\,>r_g$, where $r_g\,\sim\,\frac{GM}{c^{2}}$, is the Schwarzchild radius of the central blackhole. If $\psi$ is the semi aperture angle of the conical jet, then the location of emission region from the central black hole would be $r \sim R/\psi$. 
 The minute timescale variability suggest a compact emission region situated at sub-parsec scales from within the broad line region \cite{1994ApJS...90..945D,1995ApJ...441...79B}. Conversely, for the $\gamma$-rays to escape from the pair production losses with the ambient photon field (BLR), the emission region should be sufficiently far from the central engine \cite{Dondi_1995}. 

Modelling the lightcurve variability patterns as well as the spectral energy distribution (SED) are powerful tools for deciphering the underlying particle distribution and the physical processes responsible for the broadband emission. The multi-wavelength SED of 4C\,31.03 during the quiescent phase has been previously modeled under hadronic scenario using proton synchrotron emission \cite{2020ApJ...893L..20L}. Due to the decade long quiescent phase, this source has been sparsely studied. 
In this paper, we performed a comprehensive study on the major $\gamma$-ray outbursts of 4C\,31.03 using its variability patterns and multi-wavelength observations. As before mentioned, up to authors' knowledge, this is the first study reporting the variability and multi-wavelength aspects of these flaring events from 4C\,31.03.
The paper is organized as follows: 
in the next section, we provide the details of the selected observations and the data analysis procedures. In \S\ref{sec:dis}, we present our results and discuss the inferences drawn from our study in detail. In the last section, we conclude our findings. Throughout this work, we used the cosmology with $\Omega_m = 0.3$, $\Omega_\lambda = 0.7$ and $H_0 = 70$ km s$^{-1}$\,Mpc$^{-1}$.
\begin{figure*}
\centering
\includegraphics[scale=0.4]{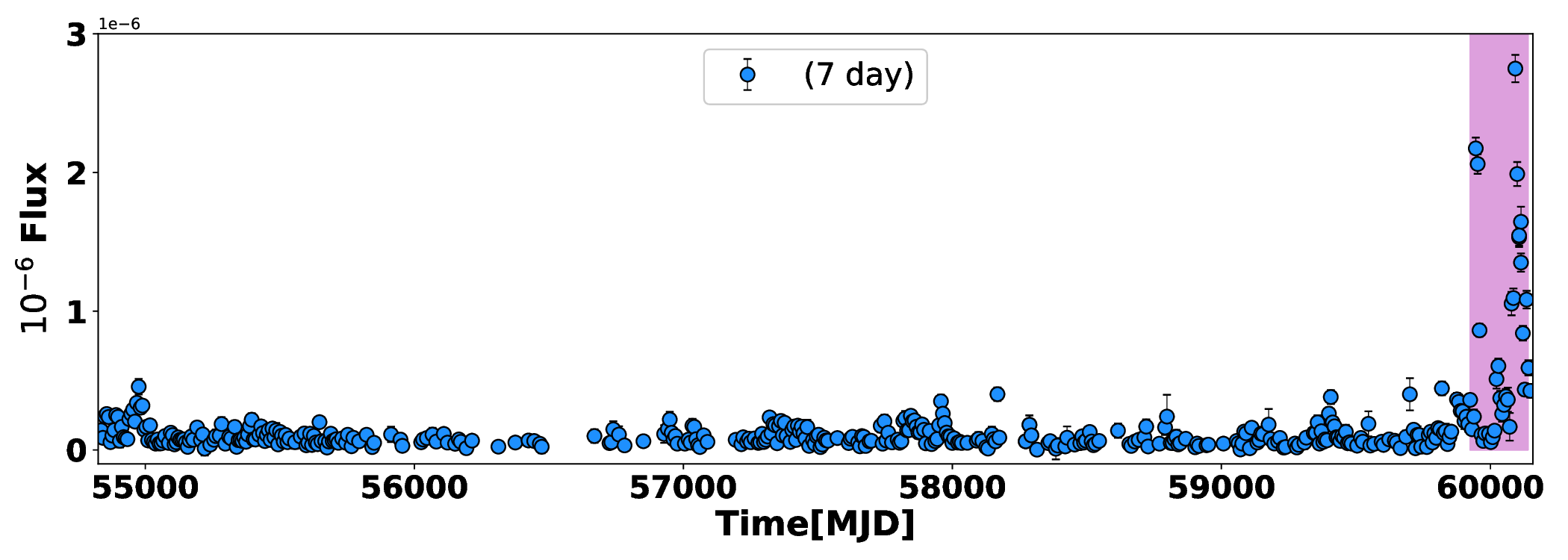}
\caption{7 day binned lightcurve of 4C\,31.03 generated using \emph{Fermi}-LAT observations spanning the time 54832-60152 MJD. The shaded time period represents the observed recent high activity epoch. The $\gamma$-ray fluxes are in the units of phs cm$^{-2}$ s$^{-1}$. }
\label{fig:lc1}
\end{figure*}
\section{Multiwavelength Observations and Data Reduction}
\label{sec:data}
\subsection{$\gamma$-ray Observations}

The \emph{Fermi}-LAT observation in the energy range 0.1-500 GeV spanning the time interval 2009 - 2023 (MJD 54830-60160) has been used in this work to study the $\gamma$-ray emission of 4C\,31.03. The region of interest (ROI) selected for this analysis was a 10 degree circular region centered at RA=18.2227 and DEC=32.1399. We used the software package \emph{Fermitools} version 2.2.0 for data reduction and analysis. The photon events were filtered and extracted the good time intervals (GTI) following standard criteria mentioned in \footnote{\url{https://fermi.gsfc.nasa.gov/ssc/data/analysis}}. The livetime cube and exposure map have been generated using \emph{`gtltcube' and `gtexpmap'} tools. A circular region of radius 25 degree centered at the target has been selected as the source region and exposure map is generated for the same. We have optimized the ROI of all the sources of interest with models from the 4FGL catalog. An initial sky model for all the sources of interest has been obtained after performing a binned likelihood analysis over the total observation. The parameters of the sources within the ROI have been kept free while that of the other sources were frozen to their catalog value. We have also applied a selection, based on the value of test statistics, TS, such that the parameters of all sources for which TS (in the entire time interval) $<9$ were kept fixed. The sky model thus acquired has been used in the preparation of lightcurves and the $\gamma$-ray SEDs. We have generated a 7 day binned global light curve of 4C\,31.03 for the entire observation. The $\gamma$-ray lightcurves during the recent active phase were obtained with two time bin sizes, 0.5 and 1 day binning. We adopted the unbinned likelihood analysis for obtaining the light curves. The PowerLaw2 function was chosen to model the $\gamma$-ray spectrum of 4C\,31.03 while generating the lightcurves. In order to obtain the fit convergence in each time bin, we made an iterative procedure: initailly, the `DRMNFB' optimizer was applied to fit in the individual time bins and whenever this fit failed to converge, we deleted all the sources with TS$<5$. In the final step, we performed the fit within all time bins after setting `Newminuit' as the optimizer. 
\subsection{\emph{Swift} Observations}
The \emph{Swift} telescope has also monitored 4C\,31.03 simultaneously with the recent outburst in the X-ray and optical/UV bands. To study the X-ray spectrum, we obtained the ready-to-use \emph{Swift}-XRT spectra from the automated online tool \emph{Swift}-XRT data products' generator \cite{2009MNRAS.397.1177E}. This tool automatically selects the source and background regions based on count rate and also performs the corrections for pile up and bad pixels. All the spectra were rebinned to 20 minimum counts in each energy bin using ftool `grppha'. Then the spectra were loaded and fitted in the X-ray fitting platform XSPEC \cite{1996ASPC..101...17A}. The unabsorbed fluxes in the selected energy bins were estimated after setting the value of hydrogen column density to $n_{H}=7.8\times10^{20}\,cm^{-2}$ \cite{Kalberla_2005}. We have also obtained the values of the photon indices and the unabsorbed integrated fluxes in the energy range 0.3-10 keV for each observation.  

The UVOT observations were downloaded from the archive and reduced following standard procedures \footnote{\url{https://swift.gsfc.nasa.gov/analysis/threads}}. The images in all the available filters were summed over the extensions using the tool \emph{uvotimsum}. A circular region of radius 6 arcsec was used to extract the source counts, while a 20 arcsec radius circle in the source free region was used for background estimation. For all observations, we also produced spectral products using the tool \emph{uvot2pha} and these were fitted in XSPEC to estimate the total integrated flux in the energy range 2-7 eV. All the fluxes in each filter as well as the integrated fluxes were corrected for galactic absorption by fixing E(B-V)=\,0.0503 for R$_v$=3.1 \cite{Schlafly_2011}.
The results of the data reduction and analysis of \emph{Swift} observations are presented in Table \ref{tab:swift}.
\begin{table*}
	\centering
	\setlength{\tabcolsep}{4pt}
	\scriptsize
	\caption{The summary of the Data analysis performed for the \emph{\emph{Swift}} XRT and UVOT observations.}
	\label{tab:swift}
	\begin{tabular}{llccccr}
		\hline
		\hline
		&&&\emph{Swift}-XRT& & &\\
		\hline
		state && Flux (ergs cm$^{-2}$ s$^{-1}$) & Photon Index ($\Gamma$/$\alpha$)& Curvature ($\beta$) & $\chi^{2}$/Dof&  \\
		\hline
		Flare1 & &$(3.6 \pm 0.3) \,\times 10^{-11}$ & $2.8 \pm 0.2$ & $-0.6\pm 0.1$ & 58.5/80& \\
		Flare2 & &$(1.5\pm 0.1) \,\times 10^{-11}$ & $1.4 \pm 0.07$ & - & 45.7/49& \\
		Quiescent & &$(5.6\pm0.4) \,\times 10^{-12}$ & $1.5 \pm 0.1$ & - & 42/35& \\
		\hline
		&&&\emph{Swift}-UVOT (Fluxes in phs cm$^{-2}$ s$^{-1}$)& &&\\
		\hline
		&V &B& U& UVW1& UVM2& UVW2\\
		\hline
		
		Flare1 & 3.8 $\pm$ 0.3 & 4.8 $\pm$ 0.3 & 3.7 $\pm$ 0.3 & 3.0 $\pm$ 0.2 & 2.1 $\pm$ 0.1 & 2.6 $\pm$ 0.2\\
		
		Flare2 & 3.0 $\pm$ 0.07 & 3.7 $\pm$ 0.06 & 2.6 $\pm$ 0.05 &  1.96 $\pm$ 0.04 & 1.3
		$\pm$ 0.03 & 1.5 $\pm$ 0.02 \\
		
		Quiescent & 0.37 $\pm$ 0.01 & 0.43 $\pm$ 0.01 & 0.29 $\pm$ 0.01 & 0.23 $\pm$ 0.01 & 0.18 $\pm$ 0.01 & 0.20 $\pm$ 0.005 \\ 
		\hline
		\hline
	\end{tabular}
\end{table*}
\begin{figure*}
\centering
\includegraphics[scale=0.45]{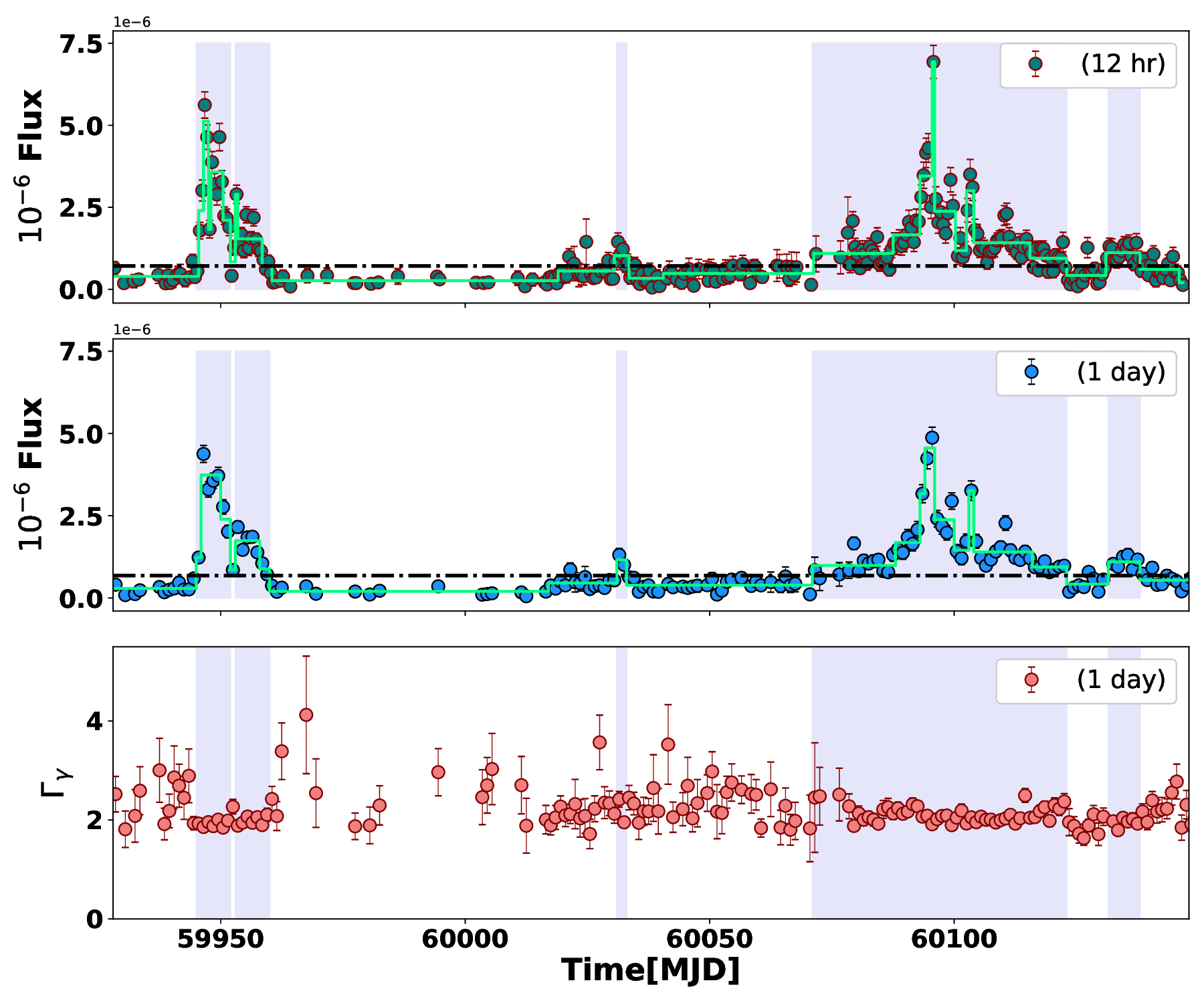}
\caption{\emph{Fermi} $\gamma$-ray lightcurve of 4C\,31.03 spanning the time 59940-60152 MJD. Top two panels show the 12 hour and 1 day binned lightcurves. Fluxes are in units of phs cm$^{2}$ s$^{-1}$. The solid line indicate the Bayesian blocks obtained while horizontal dashed line represents the average flux level. In the bottom panel, the photon index obtained from 1 day binned lightcurve has been plotted. The shaded time periods represent the duration of flares as characterized from the HOP analysis.}
\label{fig:lc2}
\end{figure*}
\section{RESULTS AND DISCUSSION}
\label{sec:dis}
\subsection{$\gamma$-ray temporal analysis of 4C\,31.03}
\label{sec:hop}
In Figure \ref{fig:lc1}, we show the 7-day binned $\gamma$-ray  lightcurve obtained for the time duration 54832-60152 MJD. It is evident from the figure that the source 4C\,31.03 has shown such pronounced activity (shaded region) for the first time since it's discovery. 
In this work, we imposed two conditions on all the lightcurves studied, which are : 1) In each time bin, test statistics, TS$\geq$ 4 and 2) uncertainty in flux, $\Delta$F $\leq$ 0.5\,F.  
There are two reported instances during which the flux exhibited an extraordinary increase \cite{2023ATel15841....1C, 2023ATel16068....1C}. In order to achieve better resolution in time and to differentiate the rise and decay substructures in this unprecedented high activity period, we obtained a 1 day as well as 12 hour binned lightcurves during the shaded region in Figure \ref{fig:lc1}. The highest flux obtained is $4.8\,\times\,10^{-6}$ phs cm$^{-2}$ s$^{-1}$
for one day binned and $6.9\,\times\, 10^{-6}$ phs cm$^{-2}$ s$^{-1}$ for 12 hour binned lightcurve. 
\begin{table}
\centering
\setlength\tabcolsep{7pt}
\caption{The duration and peak times of the flares identified.}
\label{tab:hop}
\begin{tabular}{cllr}
\hline\hline
Flare & T$_{\rm start}$ & T$_{\rm peak}$ & T$_{\rm stop}$ \\
& MJD & MJD & MJD \\
\hline
1 & 59945.0 & 59948.0 & 59952.0 \\
2 & 59953.0 & 59955.5 & 59960.0 \\
3 & 60031.0 & 60032.0  & 60033.0 \\
4 & 60071.0 &  60095.0 & 60123.0 \\
5 & 60131.5 & 60134.7 & 60138.0 \\
\hline\hline
\end{tabular}
\end{table}
To have a systematic identification of the duration of flares, we employed an algorithm which was used in previous studies \cite{2019ApJ...877...39M, 2022icrc.confE.868W}.
This algorithm utilizes Bayesian blocks and starts by identifying a block with higher flux than its preceding and succeeding blocks. It then proceeds in both directions from the peak, as long as sequentially lower blocks are found. The start/end of a flare is set by the flux exceeding/falling under a predefined flux level (average flux). This approach is referred to as the "HOP algorithm," and was first introduced by \cite{1998ApJ...498..137E}. We identified five epochs of flaring in the 12 hour and 1 day binned lightcurves shown as the shaded regions in Figure \ref{fig:lc2}. 
The top 2 panels display the 12 hour and one day binned lightcurves along with the Bayesian blocks respectively. In the bottom panel, we plotted the $\gamma$-ray photon index, $\Gamma_{\gamma}$, obtained from the 1 day binned lightcurve. It is apparent from the figure that $\gamma$-ray spectrum shows a clear hardness during the flares. Out of all the identified flares, 3 time intervals (flares 1,2, and 4) corresponding to MJD 59945-59952, 59953-59960, and 60071-60123, show prominent activity. The other 2 epochs exhibit less significant activity, especially flare 3. The time periods corresponding to the beginning and the decay of flares as well as the period of maximum flux, are tabulated in Table \ref{tab:hop}. The Bayesian blocks and the HOP groups in this analysis were obtained using an open source python module provided by
\citet{2022icrc.confE.868W}. Furthermore, the variation of $\Gamma_{\gamma}$ around 2 during the flares hints that the external Compton peak falls in the \emph{Fermi} energy range ($>$ 100 MeV). 

\subsubsection{Searching for the fastest $\gamma$-ray variability}
\label{sec:tvar}
To investigate the fastest flux doubling or halving time during the flares, we scanned the 12 hour binned lightcurve using the equation:
\begin{equation}
\centering
F(t) = F(t_0)\,2^{{-(t-t_0)}/\tau}
\end{equation}
where $F(t)$ and $F{(t_0)}$ are the fluxes at consecutive times $t$ and $t_0$ respectively and $\tau$ is the characteristic time scale.
We further imposed a minimum 3$\sigma$ significance level for the difference in flux at consecutive times \cite{2013A&A...555A.138F} for this analysis. The values of $\tau$ corresponding to the times at which the above condition has been satisfied are shown in Table \ref{tab:tvar}. Our analysis shows variability time scales ranging from 5 to 18 hours, with the fastest flux decay between MJD 59951-59952 in a time scale of $5.5\pm0.7$ hours. Following the light travel arguments discussed in \S\ref{sec:intro}, the size of the emission region, R for a $\tau$ in the order of hours, will be roughly in the range $10^{15}-10^{16}$ cm \cite{Ghisellini_2009,2009ApJ...692...32D}. 
An observed hour scale flux variability time usually indicates further shorter cooling time scale for the emitting particle distribution. And it is worth mentioning that a cooling time scale lesser than one hour supports the BLR while more than 10 hour suggests the molecular torus as the emission location in blazars \cite{2011A&A...530A..77F}. 
\begin{table*}
	\centering
	\setlength{\tabcolsep}{11pt}
	\scriptsize
	\caption{The summary of the variability times found. t$_0$ and t$_1$ are the times in MJD; F(t$_0$) and F(t$_1$) are the fluxes in units of ($\times$10$^{-6}$) phs cm$^{-2}$ s$^{-1}$; significance of difference in flux in $\sigma$; variability time scale ($\tau$); uncertainty in $\tau$; Rise or Decay.}
	\label{tab:tvar}
	\begin{tabular}{lllllllr}
		\hline
		\hline 
		t$_0$ & t$_1$ & F(t$_0$) & F(t$_1$) & Significance & $\tau$ (hour) & $\Delta\tau$ (hour) & R/D \\
		\hline\hline
59945.25 	 & 59945.75 	 & $ 0.6 \pm 0.2 $ 	 & $ 1.8 \pm 0.3 $ 	 & 3.8 	 & 7.2 	 & 1  & R\\
59945.75 	 & 59946.25 	 & $ 1.8 \pm 0.2 $ 	 & $ 3 \pm 0.3 $ 	 & 3.0 	 & 16 	 & 0.6 & R \\
59946.25 	 & 59946.75 	 & $ 3 \pm 0.3 $ 	 & $ 5.6 \pm 0.4 $ 	 & 5.1 	 & 13 	 & 0.5& R \\
59947.25 	 & 59947.75 	 & $ 4.6 \pm 0.4 $ 	 & $ 1.8 \pm 0.3 $ 	 & 6.1 	 & 9 	 & 0.6 &D \\
59947.75 	 & 59948.25 	 & $ 1.8 \pm 0.3 $ 	 & $ 3.9 \pm 0.3 $ 	 & 4.9 	 & 11 	 & 0.6 & R \\
59949.25 	 & 59949.75 	 & $ 2.9 \pm 0.3 $ 	 & $ 4.6 \pm 0.4 $ 	 & 3.3 	 & 17.6 	 & 0.5& R \\
59951.75 	 & 59952.25 	 & $ 1.9 \pm 0.2 $ 	 & $ 0.41 \pm 0.08 $ 	 & 4.7 	 & 5.5 	 & 0.7 &D \\
59952.75 	 & 59953.25 	 & $ 1.3 \pm 0.2 $ 	 & $ 3 \pm 0.3 $ 	 & 4.5 	 & 10	 & 0.7 & R\\
59953.25 	 & 59953.75 	 & $ 2.9 \pm 0.3 $ 	 & $ 1.4 \pm 0.2 $ 	 & 4.3 	 & 11 	 & 0.7  & D\\
59954.75 	 & 59955.25 	 & $ 1.2 \pm 0.2 $ 	 & $ 2.3 \pm 0.2 $ 	 & 3.4 	 & 13 	 & 0.7 & R \\
59955.25 	 & 59955.75 	 & $ 2.3 \pm 0.2 $ 	 & $ 1.3 \pm 0.2 $ 	 & 3.0 	 & 14 	 & 0.8 & D\\
60078.75 	 & 60079.25 	 & $ 0.8 \pm 0.2 $ 	 & $ 2.1 \pm 0.3 $ 	 & 3.6 	 & 8.2 	 & 1 & R\\
60094.75 	 & 60095.25 	 & $ 4.3 \pm 0.4 $ 	 & $ 2.5 \pm 0.3 $ 	 & 3.2 	 & 15.4 	 & 0.6 & D\\
60095.25 	 & 60095.75 	 & $ 2.5 \pm 0.3 $ 	 & $ 7 \pm 0.5 $ 	 & 7.2 	 & 8.2	 & 0.6 & R\\
60095.75 	 & 60096.25 	 & $ 7 \pm 0.50 $ 	 & $ 2.8 \pm 0.4 $ 	 & 6.7 	 & 9 	 & 0.5 & D\\
60099.75 	 & 60100.75 	 & $ 2.5 \pm 0.3 $ 	 & $ 1 \pm 0.2 $ 	 & 3.7 	 & 18 	 & 2 & D\\
		\hline
		\hline
	\end{tabular}
\end{table*}
\begin{figure*}
	\begin{subfigure}{0.45\textwidth}
		\includegraphics[width=0.8\textwidth]{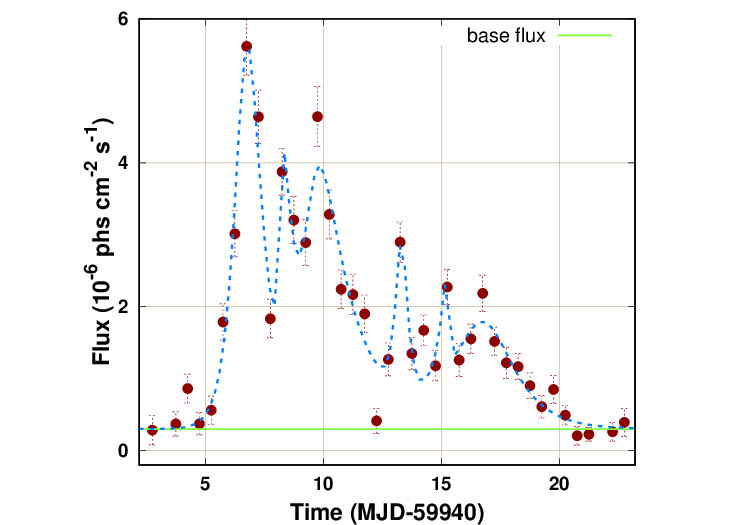}
		\label{fig:a}
	\end{subfigure}
	\begin{subfigure}{0.45\textwidth}
		\includegraphics[width=0.8\textwidth]{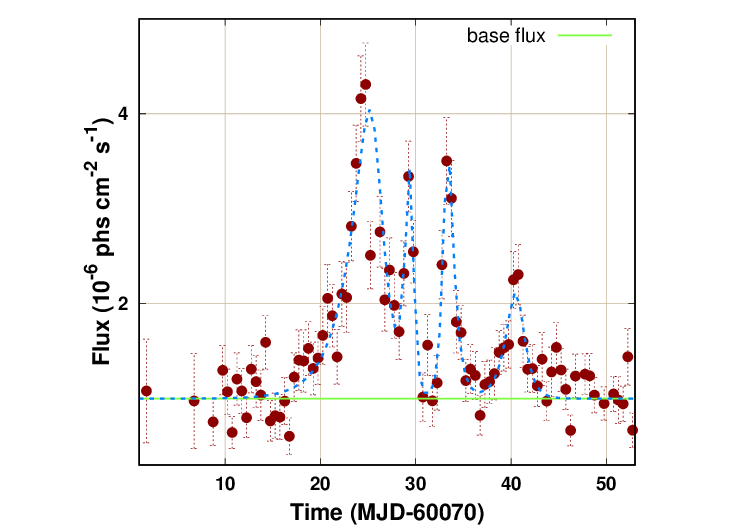}
		\label{fig:b}
	\end{subfigure}
	
	\caption{Plots showing the temporal profile fitting of the 3 prominent flares observed. The dashed curve represent the total best fit model and solid line is the base flux level chosen. Flare 1 and 2 fitted together (left). Flare 4 is plotted in the right panel.}
	\label{fig:fit}
\end{figure*}
Additionally, we performed a $\chi^{2}$ minimization fitting of the rise and decay subflares using exponential profiles represented by
\begin{equation}\
\label{eq:var2}
 F(t) = 2F_0 \left[ \exp \left( \frac{t_0 - t}{T_r} \right) + \exp \left( \frac{t - t_0}{T_d} \right) \right]^{-1}
\end{equation} 
where $T_r$ and $T_d$ are the rise and decay times of subflares, respectively, and $F_0$ is the approximated amplitude of each subflare, measured at time $t_0$. 
We considered the epochs 59945-59960 and 60071-60123 including the three prominent flares 1,2 and 4 as given in Table \ref{tab:hop}.
All the fluxes were estimated above a constant base flux in the analysis. The plots of the flares with the total best fit function are shown in Figure \ref{fig:fit}. 
Certain sub-flares within the flare 1 and 2 contain only three data points and cannot be fitted with the equation \ref{eq:var2}. Hence the parameters of these 
subflares (2,4, and 5) were kept fixed at their optimized values. These fixed parameters are given in Table \ref{tab:fit} with suffix $\ast$. For the sub-flares
with more than three data points we quote the standard errors. We also found the 
value of asymmetry $\zeta$ for each subflare from the expression, $(T_r\,-\,T_d)/(T_r\,+\,T_d)$. The flares with $\zeta<0$ are known as fast rise and exponential decay (FRED) flares which are common in $\gamma$-ray lightcurves. Such flares support the scenario of an injection time scale faster than the radiative time scale \cite{2019ApJ...877...39M}. However, in our analysis, very few sub-flares are found with $\zeta<0$. Moreover, most of the sub-flares are slightly/moderately asymmetric since $|\zeta|< 0.7$ \cite{Abdo_2010}.
\begin{table*}
	\centering
	\setlength{\tabcolsep}{12pt}
	\caption{The rise and decay times estimated from Equation \ref{eq:var2} for all the subflares. The peak flux F$_0$ is in units of \mbox{($\times$10$^{-6}$) phs cm$^{-2}$ s$^{-1}$.}} 
	\label{tab:fit}
	\scriptsize
	\begin{tabular}{llllr}
		\hline
		\hline 
		$t_0$ & $F_0$ & $T_r$ & $T_d$ & $\zeta$ \\
		(MJD) & & (day) & (day) & \\
		\hline 
		Flare 1\& 2: & & $\chi^{2}_{\rm red}$=1.21  & dof=28& \\
		&&&&\\
		\hline
		59946.75 	 & 	 $ 5.2 \pm 0.4 $ 	 & 	 $ 0.43 \pm 0.04 $ 	 & 	 $ 0.59 \pm 0.06 $ 	 & 	 0.17 \\
		59948.20	&	$2.3^{\ast}$	&  $0.1^{\ast}$ &  $0.5^{\ast}$  & $-0.7$    \\
		59949.50 	 & 	 $ 2.9 \pm 0.3 $ 	 & 	 $ 0.37 \pm 0.08 $ 	 & 	 $ 1.4 \pm 0.2 $ 	 & 	 0.59 \\
		59953.25 	 & 	 $ 2.1^{\ast} $ 	 & 	 $ 0.21^{\ast} $ 	 & 	 $ 0.22^{\ast} $ 	 & 	 $0.02$ \\
		59955.50	&	$1.0^{\ast}$	&	$0.34^{\ast}$	&	$0.06^{\ast}$	& $0.7$ \\ 			
		59956.75 	 & 	 $ 1.5 \pm 0.2 $ 	 & 	 $ 1.2 \pm 0.3 $ 	 & 	 $ 1.2 \pm 0.2 $ 	 & 	 0.001 \\
		\hline 
		Flare 4: && $\chi^{2}_{\rm red}$=1.8 & dof=48  & \\
		&&&&\\
		\hline
		60095.75 	 & 	 $ 3.8 \pm 0.3 $ 	 & 	 $ 2.2 \pm 0.3 $ 	 & 	 $ 0.91 \pm 0.2 $ 	 & 	 -0.42 \\
		60099.50	 & 	 $ 2.2 \pm 0.5 $ 	 & 	 $ 0.55 \pm 0.2 $ 	 & 	 $ 0.30 \pm 0.1 $ 	 & 	 -0.31 \\
		60103.25 	 & 	 $ 2.5 \pm 0.5 $ 	 & 	 $ 0.40 \pm 0.1 $ 	 & 	 $ 0.64 \pm 0.1 $ 	 & 	 0.26 \\
		60110.50 	 & 	 $ 1.1 \pm 0.3 $ 	 & 	 $ 1.0 \pm 0.3 $ 	 & 	 $ 0.75 \pm 0.3 $ 	 & 	 -0.14 \\
		\hline
		\hline
	\end{tabular}
\end{table*}
\begin{figure*}
	\begin{subfigure}{0.49\textwidth}
	\vspace{3mm}
	\hspace{1.5cm}
		\includegraphics[width=0.85\textwidth]{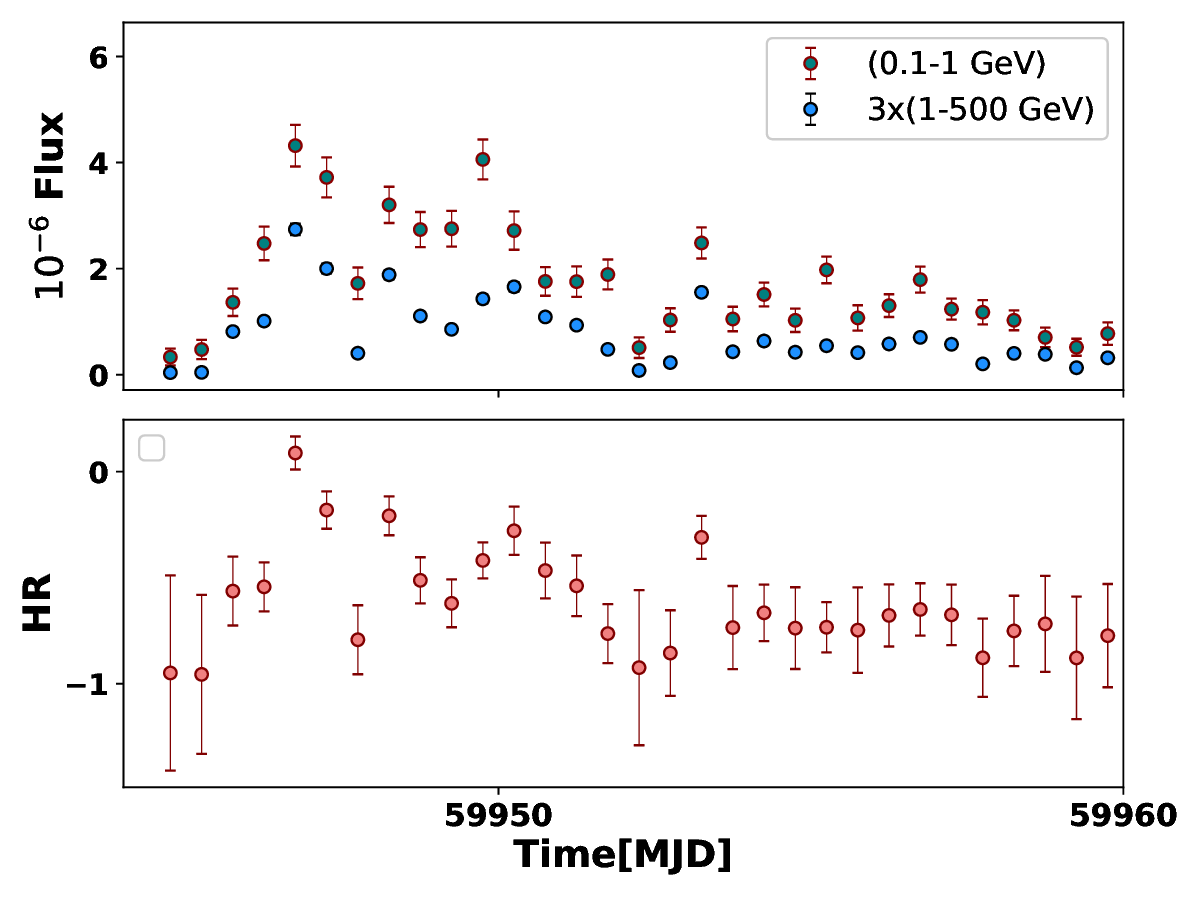}
		\label{fig:a}
	\end{subfigure}
	\hfill
	\begin{subfigure}{0.49\textwidth}
		\includegraphics[width=0.85\textwidth]{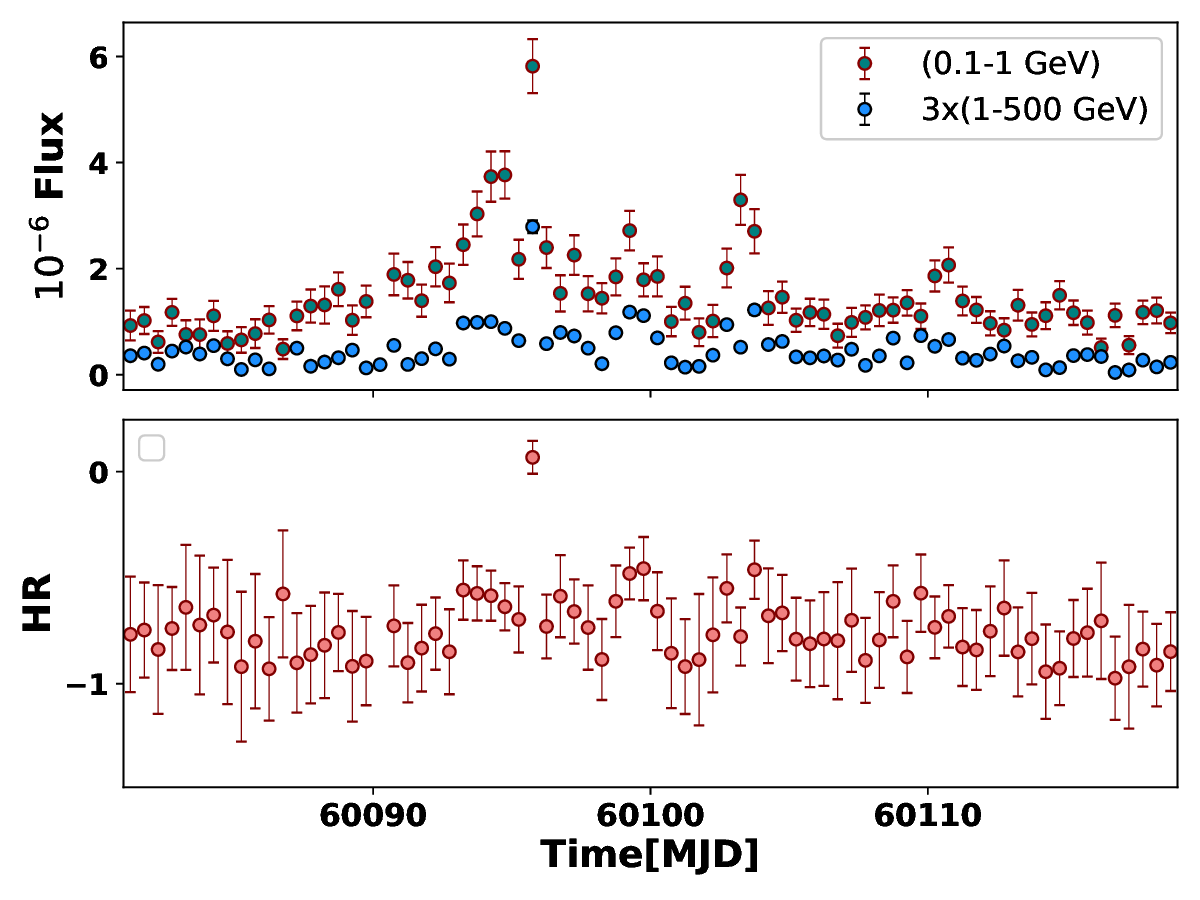}
		\label{fig:b}
	\end{subfigure}
\caption{The $\gamma$-ray ligtcurves with 12 hour binning in the low (0.1-1 GeV) and high-energy (1-500 GeV) ranges  are shown in the top panels. The high-energy light curve has been scaled to plot together. In the bottom panel, hardness ratio $HR$ has been plotted against time. Left) Flare 1 and 2 combined. Right) Flare 4.}
\label{fig:hr}
\end{figure*}
\begin{figure*}
	\begin{subfigure}{0.49\textwidth}
	\vspace{3mm}
	\hspace{1.5cm}
		\includegraphics[width=0.7\textwidth]{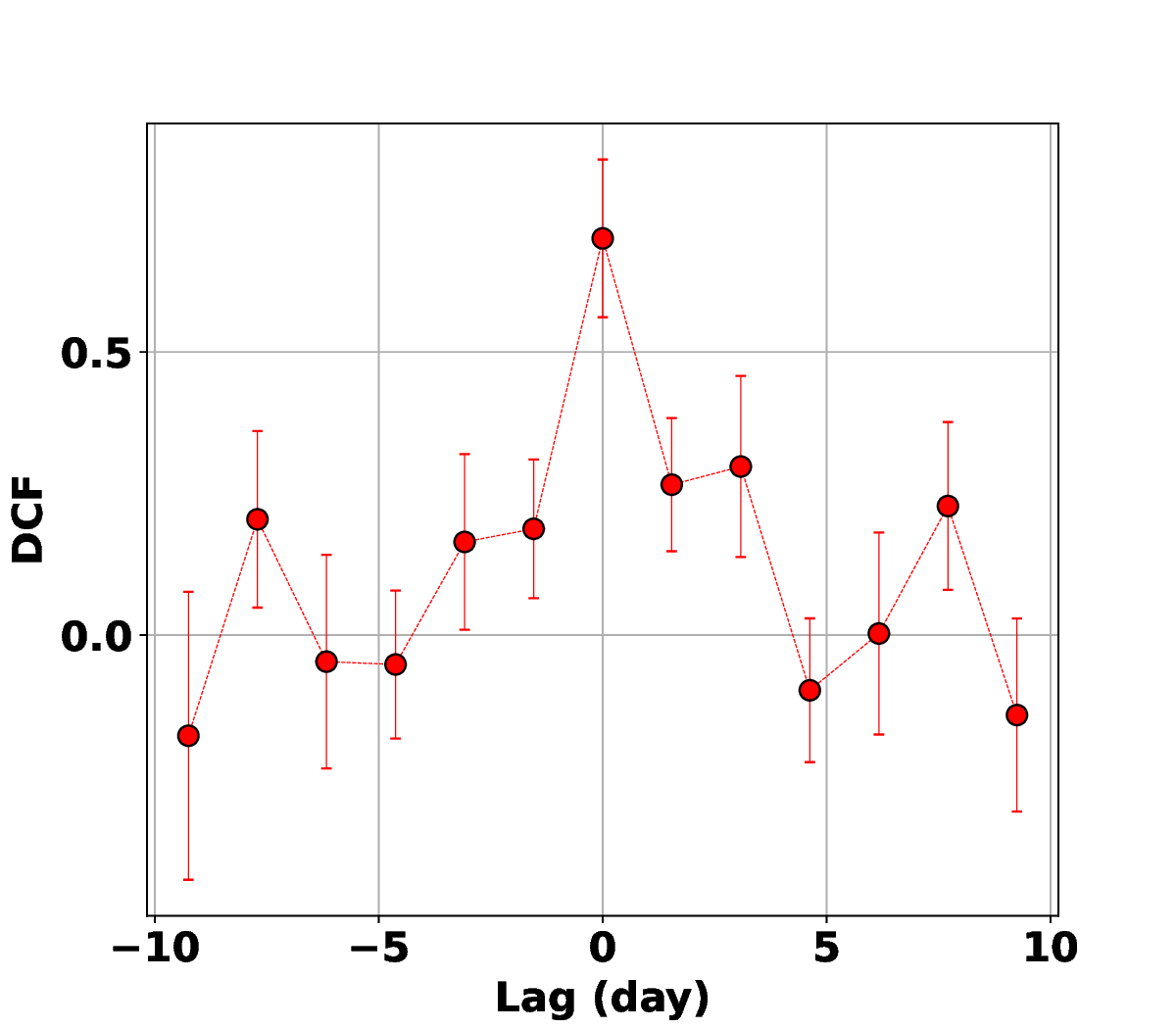}
		\label{fig:a}
	\end{subfigure}
	\hspace{-1cm}
	\begin{subfigure}{0.49\textwidth}
		\includegraphics[width=0.7\textwidth]{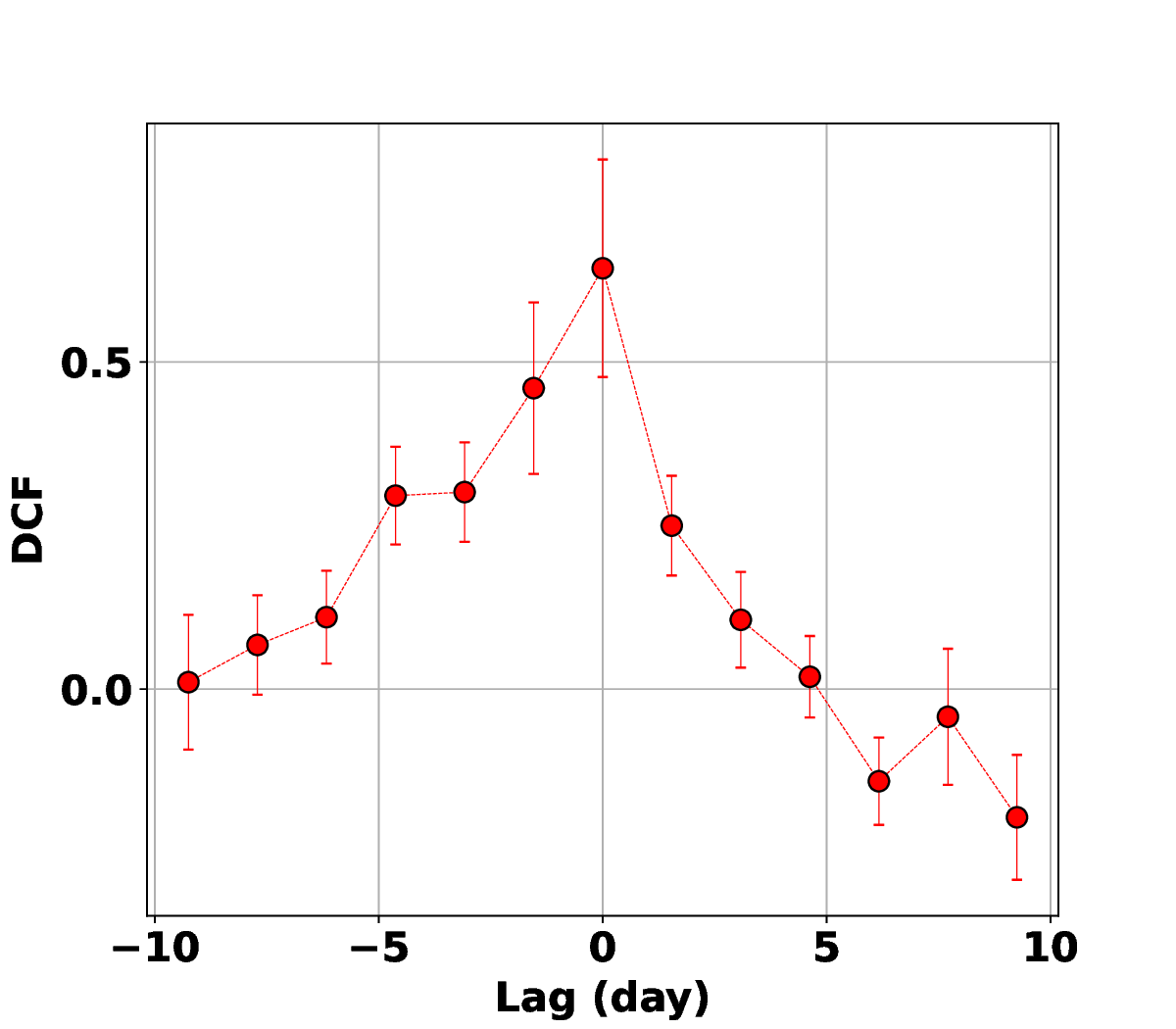}
		\label{fig:b}
	\end{subfigure}
\caption{DCF is plotted across the lag in days measured as Time\,(100-500 GeV)\,-\,Time\,(0.1-100 GeV) lightcurves. Left) DCF corresponding to flare 1 and 2 combined. Right) DCF obtained during flare 4.}
\label{fig:dcf}
\end{figure*}
\subsubsection{High and low-energy $\gamma$-ray lightcurves}
The good test statistics for the fluxes during the 3 prominent flares made it possible to split the lightcurve into  low-energy (0.1-1 GeV) and high-energy (1-500 GeV) divisions.
In Figure \ref{fig:hr}, we show the lightcurves in these energy ranges plotted together (scaled appropriately). In the bottom panel, we plotted the hardness ratio (HR) defined by, $$HR = \dfrac{f_H\,-\,f_L}{f_H\,+\,f_L}$$ where $f_H$ and $f_L$ are the fluxes in the high and low energy bands respectively.
The nature of variations in the low and high energy scenarios seems similar, however, during most of the epochs, the flaring is more intense in the low energy regime. This observation indicates that the recent active phase of 4C\,31.03 is primarily caused by the low energy electrons in the particle spectrum. 
We also explored the time lag between the low and high energy $\gamma$-ray lightcurves using the Discrete Correlation Function (DCF) \cite{Edelson_1988}. 
A zero time-lag between these lightcurves is obtained and this may suggest an energy-independent evolution of the light curves or the underlying electron distribution. 
However, the cooling timescales of the emitting electron distribution at these photon energies are too short than the 12 hour binned light curve used for the 
DCF analysis. For instance, the cooling time scale (equation \ref{eq:coolt} from \S\ref{sec:bbsed}) corresponding to $\sim$ 0.5 GeV emission is $\sim$ 7.3 hours
and the in case of $\sim$ 20 GeV it is $\sim$ 1 hour. Hence, the zero time-lag obtained through the DCF analysis indicates its inability to probe the energy-dependence
of the lightcurves obtained with 12 hour binning.


\subsubsection{$\gamma$-ray spectral variations}
\label{sec:gamma}
\begin{figure*}
 \centering
	\begin{subfigure}{0.33\textwidth}
		\includegraphics[scale=0.3]{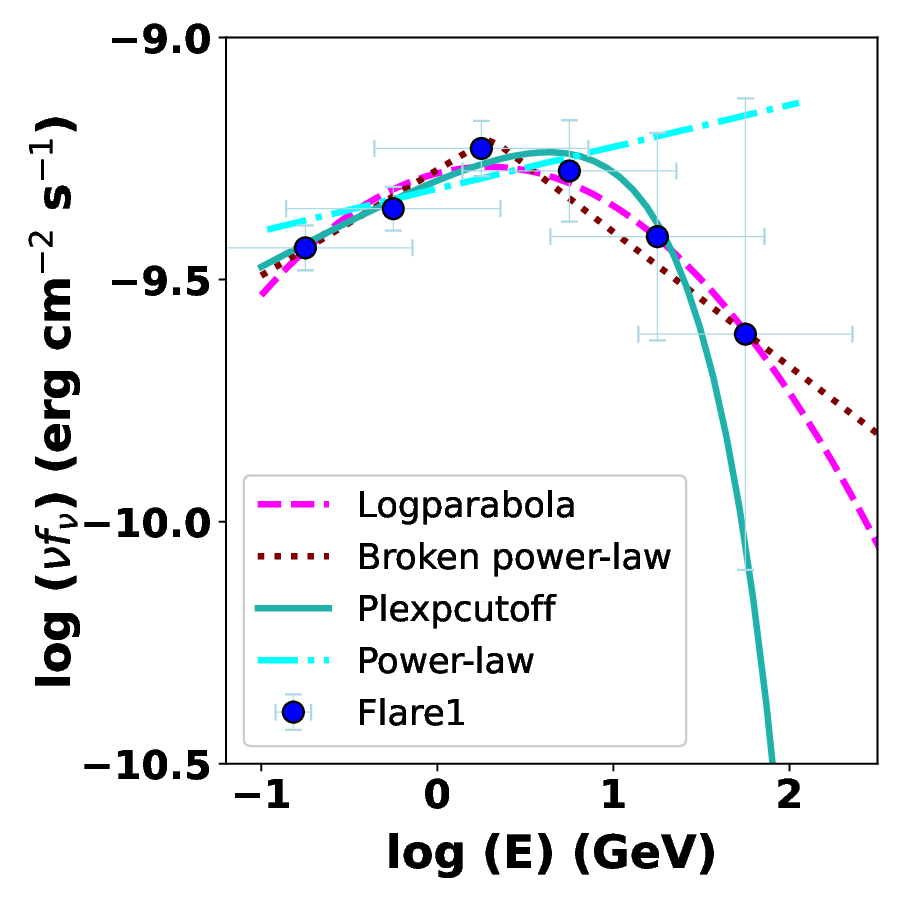}
		\caption{Flare 1}
		\label{fig:a}
	\end{subfigure}
	\hspace{-2mm}
	\begin{subfigure}{0.33\textwidth}
		\includegraphics[scale=0.3]{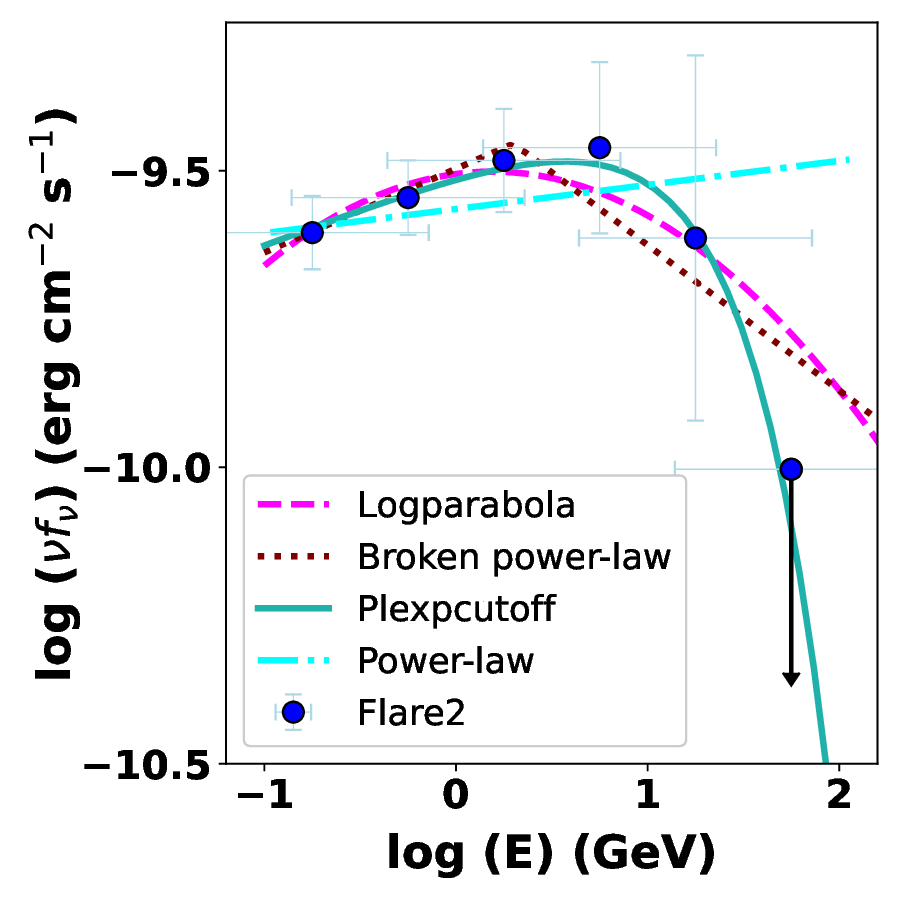}
		\caption{Flare 2}
		\label{fig:b}
	\end{subfigure}
	\hspace{-2mm}
	\begin{subfigure}{0.3\textwidth}
		\includegraphics[scale=0.3]{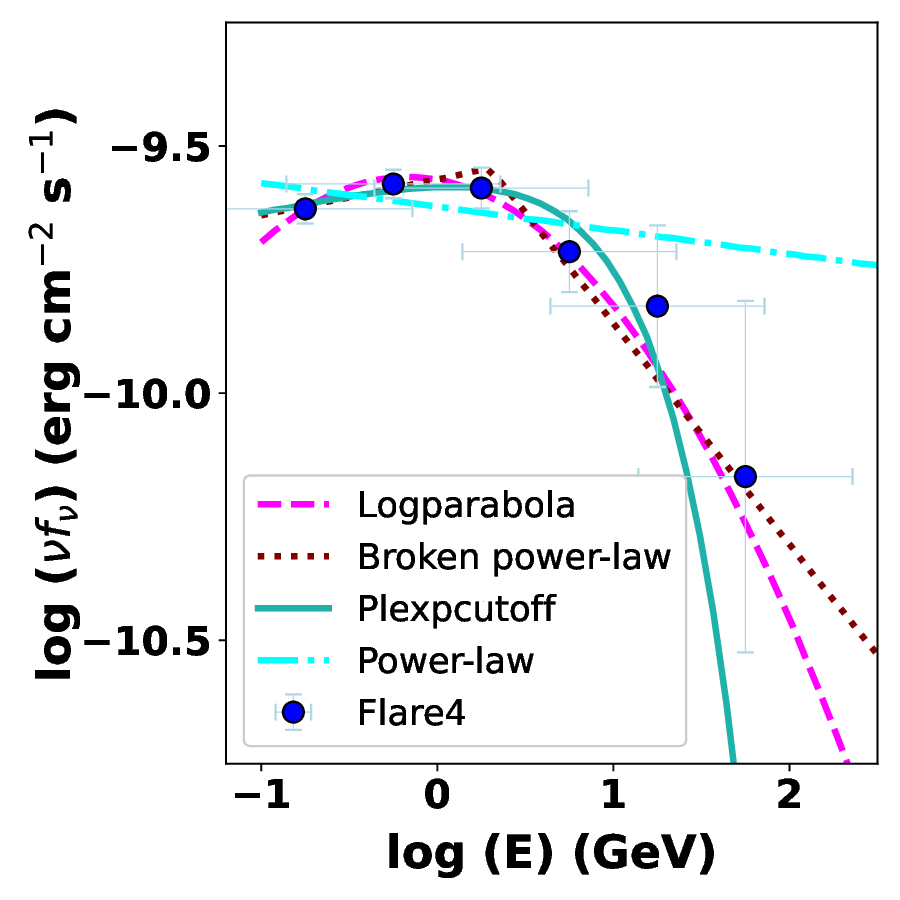}
		\caption{Flare 4}
		\label{fig:c}
	\end{subfigure}
	\smallskip
	\hfill
	\centering
	\begin{subfigure}{0.4\textwidth}
		\includegraphics[scale=0.3]{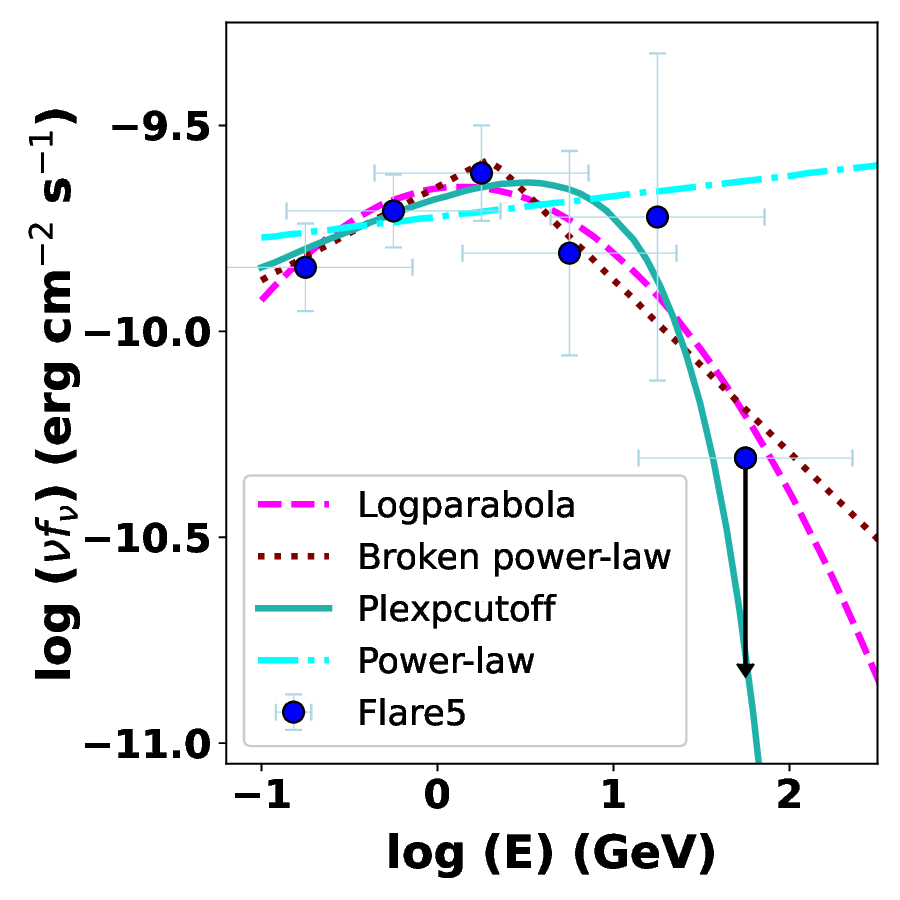}
		\caption{Flare 5}
		\label{fig:d}
	\end{subfigure}
	\hfill
	\hspace{-4cm}
	\begin{subfigure}{0.4\textwidth}
		\includegraphics[scale=0.3]{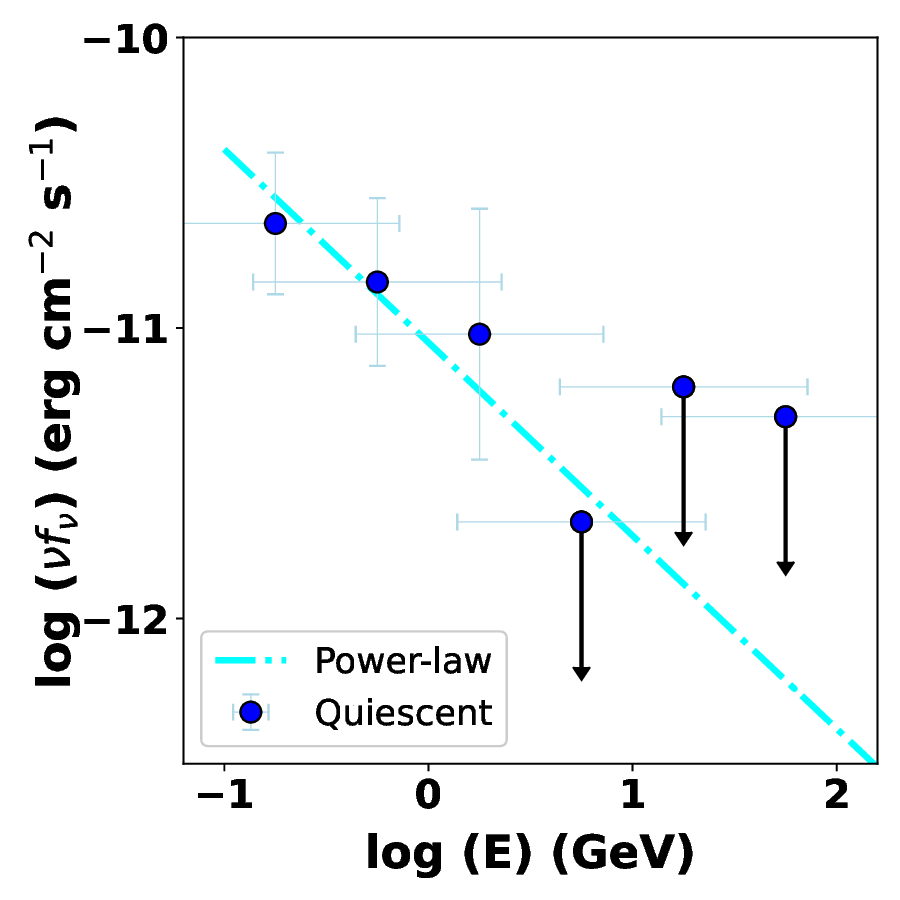}
		\caption{Quiescent}
		\label{fig:e}
	\end{subfigure}
	\hfill
\caption{Plots showing the $\gamma$-ray SEDs during different activity states along with the fitted spectral models. The details of models are inboxed in the plots.}
\label{fig:spec}
\end{figure*}

In order to study the $\gamma$-ray spectral transitions during the recent activity phase of 4C\,31.03, we obtained the corresponding  SEDs for the flares 1,2,4, and 5 (\S\ref{sec:hop}) using the \emph{likeSED.py} module. The SED for flare 3 cannot be constrained due to low photon statistics. We also extracted the $\gamma$-ray SED during the epoch 59960-59975, when the source is at a quiescent state. This SED is used later to compare the broadband SEDs during different activity states of the source (\S\ref{sec:bbsed}).
The SEDs corresponding to these epochs were fitted using power-law (PL), log-parabola (LP), broken power-law (BPL), and power-law with an exponential cutoff (PLEC) models. The PL model was able to give an acceptable fit only during the quiescent state. 
The significance of curvature/break, ($TS_{\rm curv}$) of a spectrum can be estimated as \mbox{TS (model of curvature/break) - TS (Power-law)}. According to this criterion, a spectrum shows significant curvature/break if $TS_{\rm curv}\geq 16$, as it indicates a 4$\sigma$ level confidence \cite{Prince_2020,2021MNRAS.502.5245P}.
The results of this analysis with the TS$_{\rm curv}$ values obtained are given in Table \ref{tab:spec}. The $TS_{\rm curv}$ results show that nearly all the flaring $\gamma$-ray spectra (flares 1, 2, and 4) exhibit a notable break, as BPL model has been found to give better statistics in these cases. 
 However, in the case of flares 1 and 4, the $\gamma$-ray spectra can also be well described by LP and PLEC models in addition to BPL (Table \ref{tab:spec}). In the case of flare 5, the results show that the spectrum can be described equally well by all the models. The plots of the $\gamma$-ray SEDs obtained along with the models are shown in Figure \ref{fig:spec}.

The break in the spectra is found to be almost stationary and lies around 2 GeV. The GeV spectral break in \emph{Fermi} bright blazars might be intrinsic in nature and can be attributed to the shape of the underlying particle distribution or the target photon spectrum \cite{2021MNRAS.502.5875K,2012ApJ...753..176L}. The break/curvature in the $\gamma$-ray spectrum can also be caused by the pair absorption of GeV photons by the Ly-${\alpha}$ line emission of the BLR. This will be the case only when the emission region lies within the BLR \cite{2010ApJ...717L.118P}. However, a recent study by \citet{Costamante_2018}, investigated the external Compton scenario involving BLR photons in a sample of 106 \emph{Fermi} bright FSRQs (including 4C\,31.03) suggests that the $\gamma$-ray spectra of such sources do not exhibit significant absorption effects. Instead, $\gamma$-rays seems to be produced outside the BLR.
\subsubsection{Highest photon energy}
\label{gev}
As mentioned above, the detection of photons with energy greater than 20 GeV is pivotal in the case of FSRQ type sources since the photons of such energy detected can plausibly indicate that the $\gamma$-ray production happens beyond the BLR \cite{2006ApJ...653.1089L, Costamante_2018}.
We found the energies of the photons detected during the active state of 4C\,31.03 which are positionally consistent with the source co-ordinates using the fermitool \emph{gtsrcprob}.
Three epochs were noticed at which photons with energy greater than 50 GeV have been detected from the source with more than 99\% association probability. They are 51.3 GeV at 59946.76 MJD, 82.6 GeV at 59947.15 MJD, and 65.5 GeV at 60107.24 MJD. 
\begin{table*}
	\centering
	\setlength{\tabcolsep}{12pt}
	\caption{Results of $\gamma$-ray SED analysis for various observed states.}
	\scriptsize
	\label{tab:spec}
	\begin{tabular}{l l l l l r r }
		\hline\hline
		Flare group/& $F_{0.1-300 \rm{GeV}}$& PowerLaw & & & TS & TS$_{\rm curv}$ \\
		states& (10$^{-6}$ ph cm$^{-2}$ s$^{-1}$) &  $\Gamma$ & & & & \\
		\hline
		1       & 2.6$\pm$0.08   & -1.9$\pm$0.02 & -- & -- & 6682.16  & -- \\ 
		2       & 1.7$\pm$0.07   & -2.0$\pm$0.03 & -- & -- & 3116.97 & --\\
		
		4       & 1.6$\pm$0.03 & -2.0$\pm$0.02 & -- & -- & 12894.82 & --\\
		5       & 1.1$\pm$0.07 & -1.9$\pm$0.04 & -- & -- & 1464.99  & --\\
		Quiescent      & 0.16$\pm$0.02 & -2.7$\pm$0.2 & -- & -- & 109.53 & --\\
		
		\hline
		& & LogParabola & \\
		& & $\alpha$ & $\beta$ &  & & \\
		\hline 
		1     & 2.4$\pm$0.08 & 1.8$\pm$0.03 & 0.07$\pm$0.01 & -- & 6736.82 & 54.66 \\
		2     & 1.6$\pm$0.07 & 1.9$\pm$0.04 & 0.05$\pm$0.02 & -- & 3125.17  & 8.19 \\
		4     & 1.5$\pm$0.03 & 1.9$\pm$0.02 & 0.08$\pm$0.01 & -- & 13003.37 & 108.55 \\
		5     & 1.0$\pm$0.07 & 1.8$\pm$0.07 & 0.09$\pm$0.03 & -- & 1468.75 & 4.75 \\
		
		\hline
		& & Broken PowerLaw & & $\epsilon_{\rm break}$& &  \\
		& & $\Gamma_1$ & $\Gamma_2$ & (GeV) & & \\
		\hline
		1       & 2.5$\pm$0.08 & -1.8$\pm$0.03 & -2.3$\pm$0.07 & 2.0$\pm$0.05 & 6725.74 & 43.58 \\
		2       & 1.6$\pm$0.07 & -1.9$\pm$0.04 & -2.2$\pm$0.1 & 2.0$\pm$0.07 & 3134.50  & 17.53 \\
		4      & 1.5$\pm$0.03 & -1.9$\pm$0.02 & -2.4$\pm$0.06 & 1.8$\pm$0.2 & 12965.38 & 70.56 \\
		5       & 1.0$\pm$0.07 & -1.8$\pm$0.07 & -2.4$\pm$0.2 & 2.0$\pm$0.03 & 1478.32 & 13.41 \\
		
		\hline  
		& & PLExpCutoff & &$\epsilon_{\rm cutoff}$  & & \\
		& & $\Gamma_{\rm plec}$ & &(GeV)  && \\
		\hline
		1    & 2.5$\pm$0.08 & -1.8$\pm$0.03 & --&22.0$\pm$5.0  & 6733.48 & 51.32 \\
		2    & 1.6$\pm$0.07 & -1.9$\pm$0.03 &--& 30.0$\pm$0.8  & 3110.54  & -6.43 \\
		4    & 1.5$\pm$0.03  & -1.9$\pm$0.02 & --&16.0$\pm$3.0 & 12915.87 & 21.07 \\
		5    & 1.0$\pm$0.07  & -1.8$\pm$0.07 & -- &16.5$\pm$7.4 & 1461.22 & -3.68\\
		\hline
		\hline
		
	\end{tabular}
\end{table*}
\subsection{Constraints on the Doppler factor}
\label{sec:jetparam}
The detection of $\gamma$-rays from blazars has been used to constrain various source parameters \cite{2013A&A...555A.138F, 2010MNRAS.405L..94T}. 
The observed variability timescale ($\tau$) coupled with the light travel time arguments can effectively constrain the emission region size as
\begin{equation}
\label{eq:size}
R \leq \frac{c\,\tau}{1+z}\, \delta
\end{equation}
where $\delta$ is the jet Doppler factor.
Using this relation along with the pair production opacity of the $\gamma$-ray photons against the target X-ray photons, one can  obtain a lower limit on the jet Doppler factor \cite{Dondi_1995,Abdo_2010}.
For a $\gamma$-ray photon of energy $\epsilon$ in $m_ec^{2}$ units and the target X-ray photon flux $f_x$, the minimum jet Doppler factor can be deduced as
\begin{align}
\label{eq:delta}
 \delta\, \geq  \left[ \frac{\sigma_T d_L^2 (1+z)^2 f_x \epsilon}{4 \tau m_e c^{4}}  \right] ^{\frac{1}{6}}
\end{align}
where, $\sigma_T$ is the Thomson scattering cross section and d$_L$ is the luminosity distance. 
The highest photon energy detected during the time period of flares 1 and 2 is 82.6 GeV (\S\ref{gev}) and the X-ray flux  
obtained in the energy range 0.3-10 keV during the flaring state is 2.915$\times$10$^{-11}$ erg/cm$^2$/s. Incorporating this values with shortest variability timescale of 5.5 hours (\S\ref{sec:tvar}) in the above equation, we found the minimum jet Doppler factor during the flaring state as 17. 
Using this value in equation \ref{eq:size}, the upper-limit on the emission region size R is estimated to be $\sim 10^{16}$ cm.
For a similar value of the emission region size and the target X-ray photon flux 5.6$\times$10$^{-12}$ erg/cm$^2$/s, corresponding 
to the quiescent state, the jet Doppler was found to be 13. Here, we have used the 
highest photon energy as 26 GeV (detected during the quiescent state on 59960 MJD). 
\subsection{Broadband SED analysis}
\label{sec:bbsed}
\begin{figure}
\centering
\includegraphics[scale=0.5]{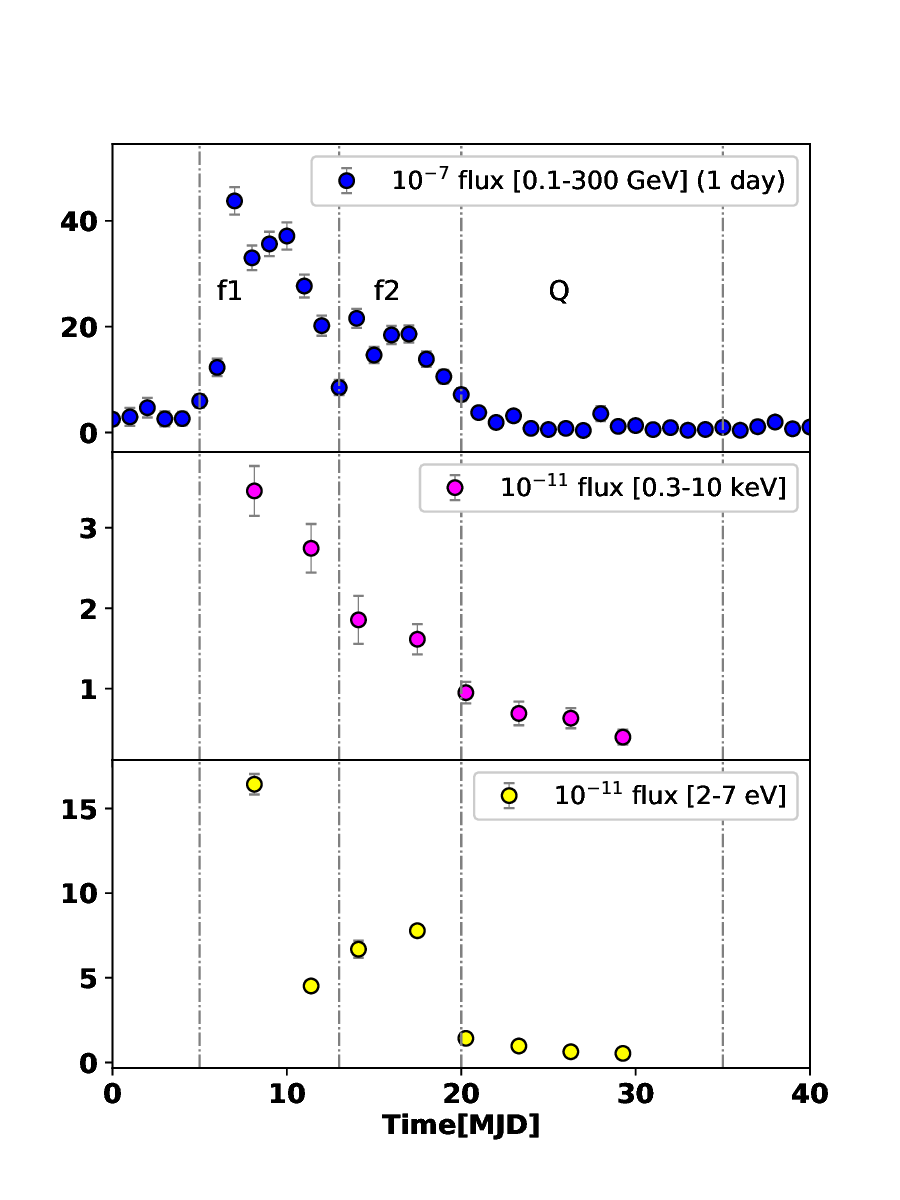}
\caption{Multi-wavelength lightcurve of 4C\,31.03 from MJD 59940-59965. Top panel shows the $\gamma$-ray flux, middle the X-ray flux and bottom panel optical/UV flux. The middle and bottom panel fluxes are in units of erg cm$^{-2}$ s$^{-1}$. $\gamma$-ray flux is in unit of phs cm$^{-2}$ s$^{-1}$.}
\label{lc-multi}
\end{figure}
\begin{figure}
\centering
\includegraphics[width=0.75\columnwidth, angle=-90]{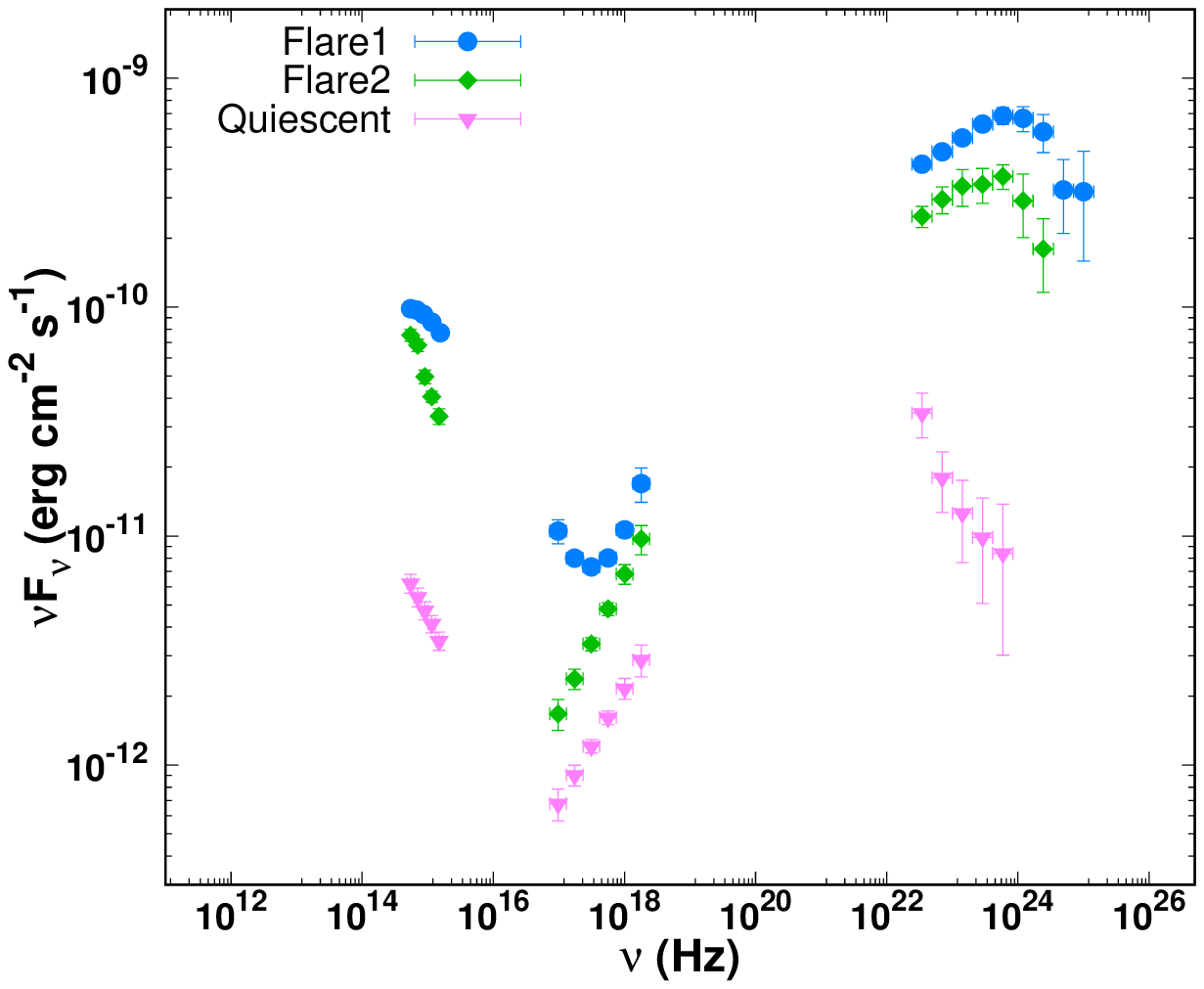}
\caption{Plot showing all the 3 multiwavelength SEDs together.}
\label{fig:sed1}
\end{figure}
The broadband spectral modeling of the source with different emission processes can provide clues on the gamma-ray emission 
mechanism and the plausible location of the emission region in the blazar jet \cite{Ghisellini_2009,Cerruti_2020}. For this,  we selected three epochs
from the multi-wavelength light curve corresponding to two high activity states and a quiescent state. These epochs were also chosen 
based on the availability of the \emph{Swift} observations and the selected time 
windows are shown in Figure \ref{lc-multi}.  In Figure \ref{fig:sed1}, we show the observed SED corresponding to these flux states.

The broadband SEDs available at optical/UV, X-ray, and $\gamma$-rays from the \emph{Swift}-XRT and \emph{Fermi} observations 
are modeled using a one-zone leptonic model considering synchrotron, SSC, and EC processes \cite{2017MNRAS.470.3283S,2018RAA....18...35S}. This model assumes 
the emission region to be a sphere of radius $R$ moving down the blazar jet with the bulk Lorentz factor $\Gamma$ and 
at an angle $\theta$ with respect to the observer. The emission region is populated with a non-thermal electron distribution
represented by a broken power-law, 
\begin{align} \label{eq:broken}
	N(\gamma)\,d\gamma = \begin{cases}
		K\,\gamma^{-p}\,d\gamma&\textrm{for}\, \mbox {~$\gamma_{\rm min}<\gamma<\gamma_b$~} \\
		K\,\gamma_b^{q-p}\gamma^{-q}\,d\gamma&\textrm{for}\, \mbox {~$\gamma_b<\gamma<\gamma_{\rm max}$~}
		\end{cases}
	\end{align}
where, $\gamma$ is the dimensionless energy of the electron and, $p$ and $q$ are the low and high-energy indices of the 
distribution with $\gamma_b$ as the break energy. 
The magnetic field responsible for the 
synchrotron emission is assumed to be tangled, and the external photon fields for the EC process are the dominant
Ly-$\alpha$ emission from the BLR and the thermal IR photons from the molecular torus. 
Due to relativistic motion, the emitted source radiation will be blue-shifted and boosted in the observer's frame 
by the Doppler factor $\delta$.

The emissivity functions corresponding to these radiative processes are solved numerically and the routines 
developed are incorporated as a local model in the spectral fitting package \emph{XSpec} \cite{1996ASPC..101...17A}. 
For numerical 
simplicity, the line emission from BLR is treated as a blackbody at a temperature of 42000K, such that the 
peak frequency is equivalent to that of the Ly-$\alpha$ emission line ($2.47 \times 10^{15} Hz$). The emission from 
the molecular torus is assumed to be a blackbody at a temperature of 1000K. 
The energy density of the target photon field involved in the external Compton scattering process is expressed
as a fraction ($f$) of the corresponding black-body energy density.
\begin{equation}
U_{\rm target}\,=f\, \times U_{\rm BB}
\end{equation} 
where, $U_{\rm BB}=\dfrac {4\sigma_B}{c} T^4 $ is the black-body energy density.
The total number of parameters
governing the broadband spectrum exceeds the information available from the optical, X-ray, and $\gamma$-ray 
energies, and hence, certain constraints were imposed. We assumed near equi-partition between the magnetic field and particle energy densities \cite{2014ApJ...782...82D,1959ApJ...129..849B}. This is defined as a parameter $\eta$ which is the ratio between particle to magnetic field energy densities.
The emission region size was initially kept free, and then optimized at $2 \times 10^{16}$ cm.  
Besides these, we also kept the parameters $\gamma_{\rm min}$=$10$, $\gamma_{\rm max}$=$2\times 10^{6}$, and viewing angle, $\theta$=$2^{\circ}$ fixed.

\begin{figure*}
\centering
	\begin{subfigure}{0.49\textwidth}
		\includegraphics[width=0.6\textwidth, angle=-90]{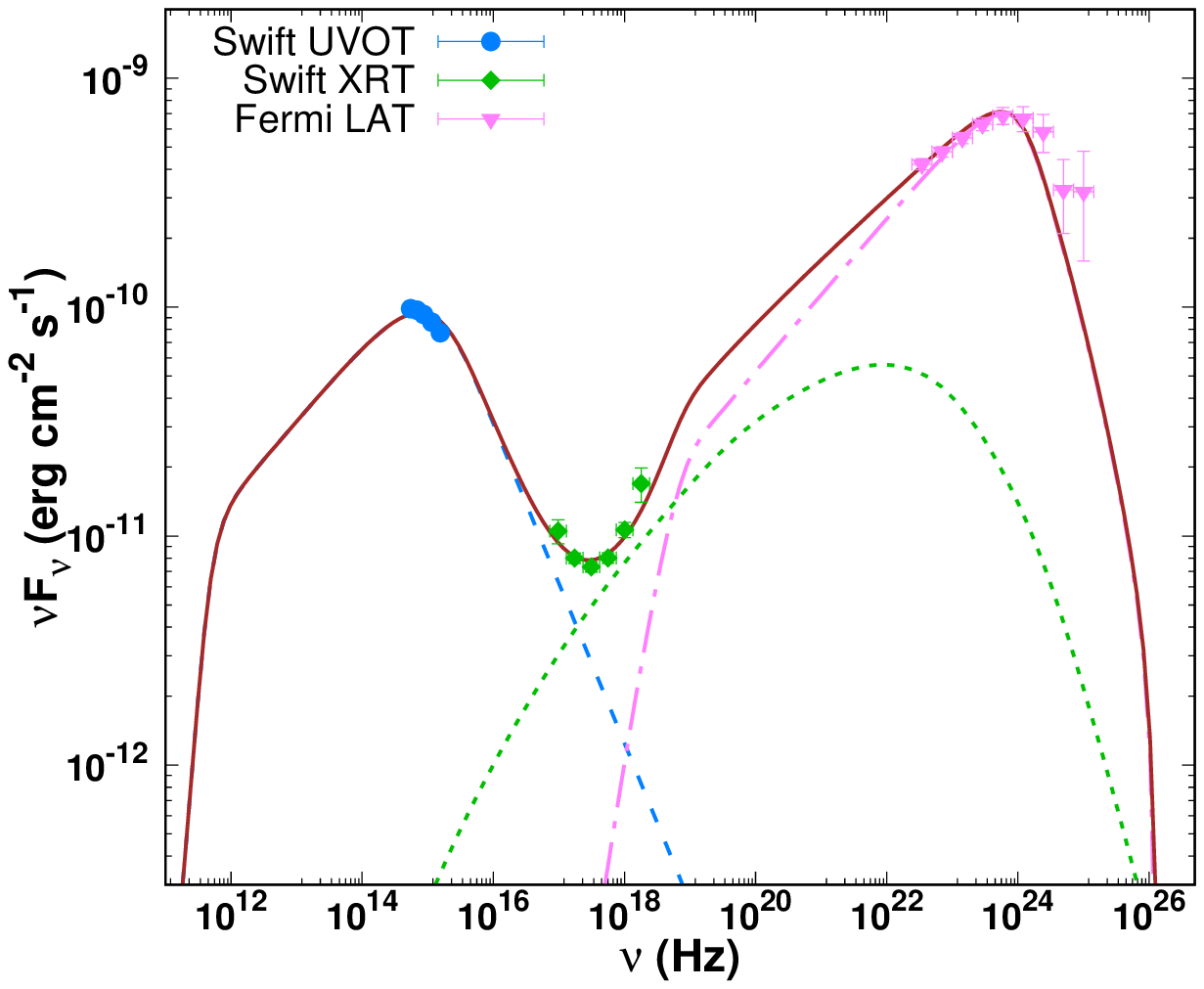}
		\caption{}
		\label{fig:a}
	\end{subfigure}
	\hspace{-2cm}
	\begin{subfigure}{0.49\textwidth}
		\includegraphics[width=0.55\textwidth, angle=-90]{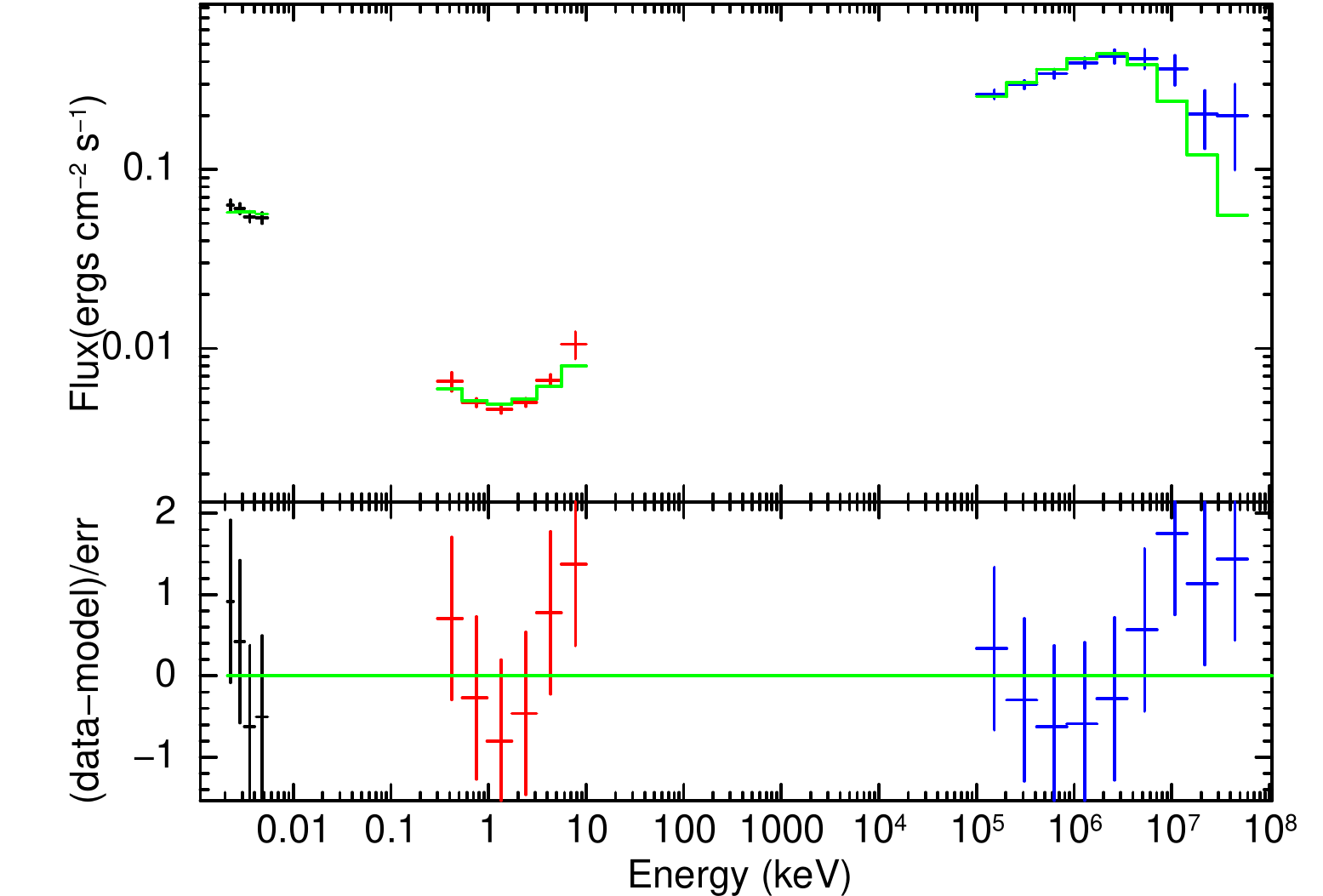}
		\caption{}
		\label{fig:b}
	\end{subfigure}
	
	\smallskip
	\begin{subfigure}{0.49\textwidth}
		\includegraphics[width=0.6\textwidth, angle=-90]{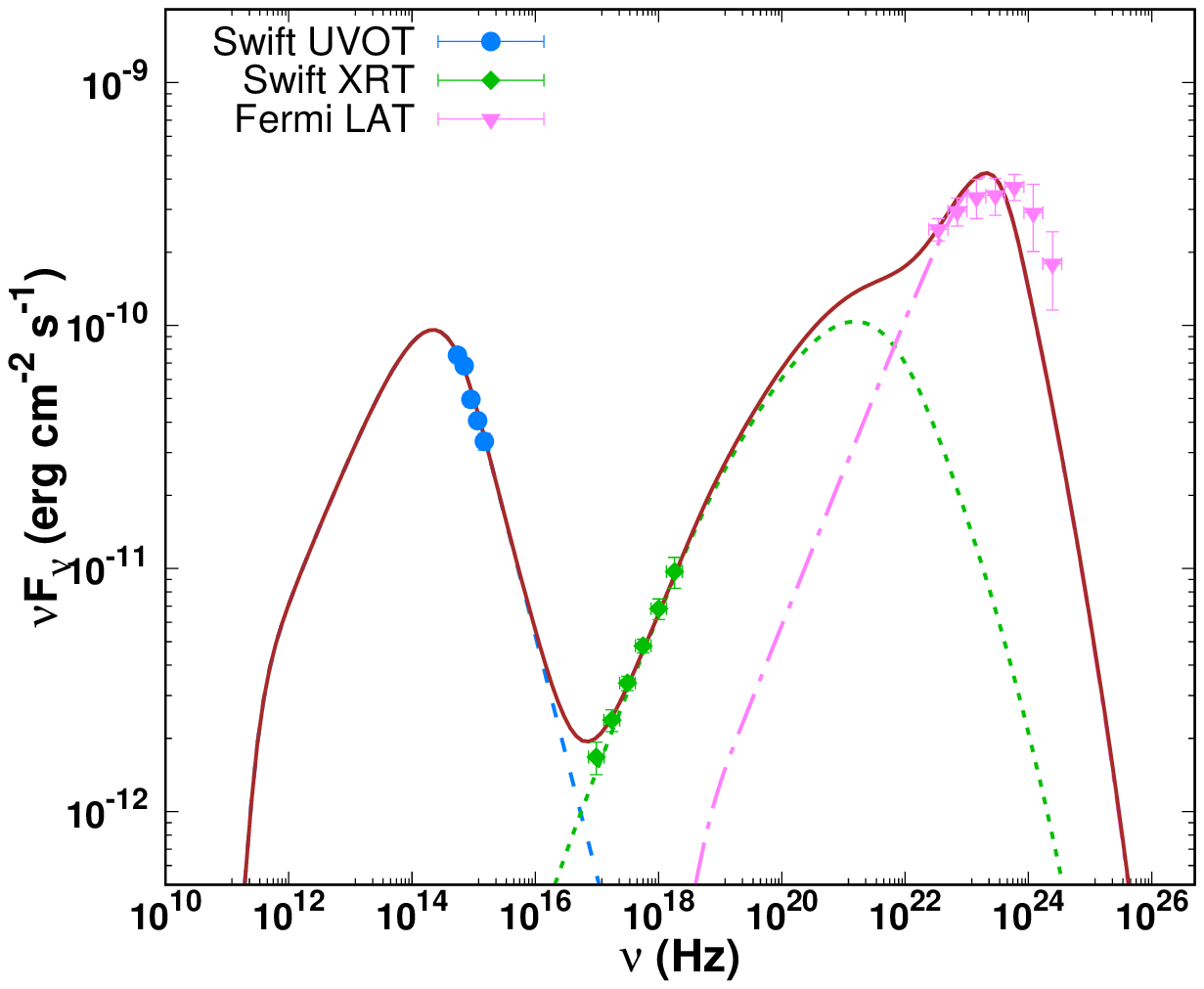}
		\caption{}
		\label{fig:c}
	\end{subfigure}
	\hspace{-2cm}
	\begin{subfigure}{0.49\textwidth}
		\includegraphics[width=0.55\textwidth, angle=-90]{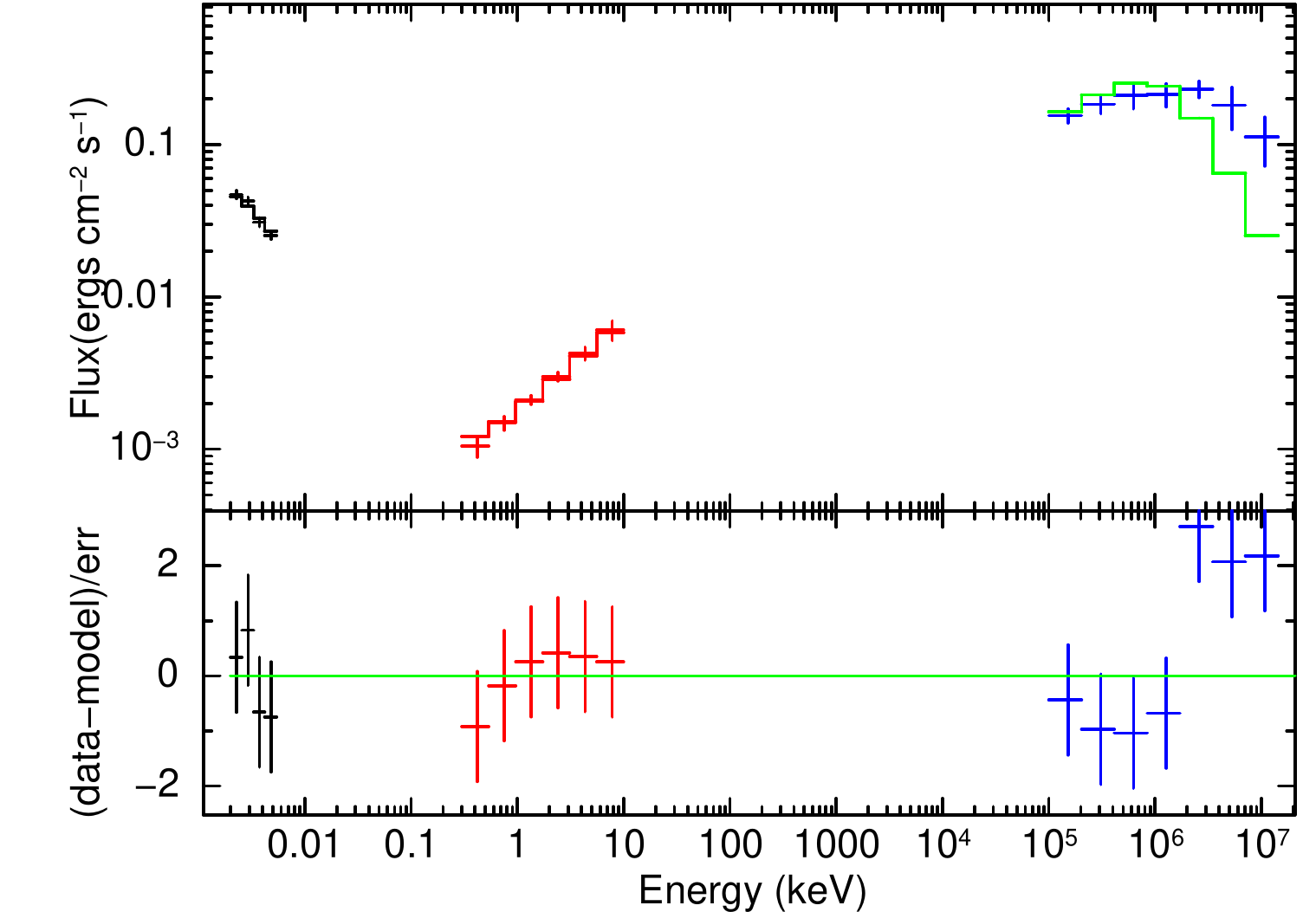}
		\caption{}
		\label{fig:d}
	\end{subfigure}
	\smallskip
	\begin{subfigure}{0.49\textwidth}
		\includegraphics[width=0.6\textwidth, angle=-90]{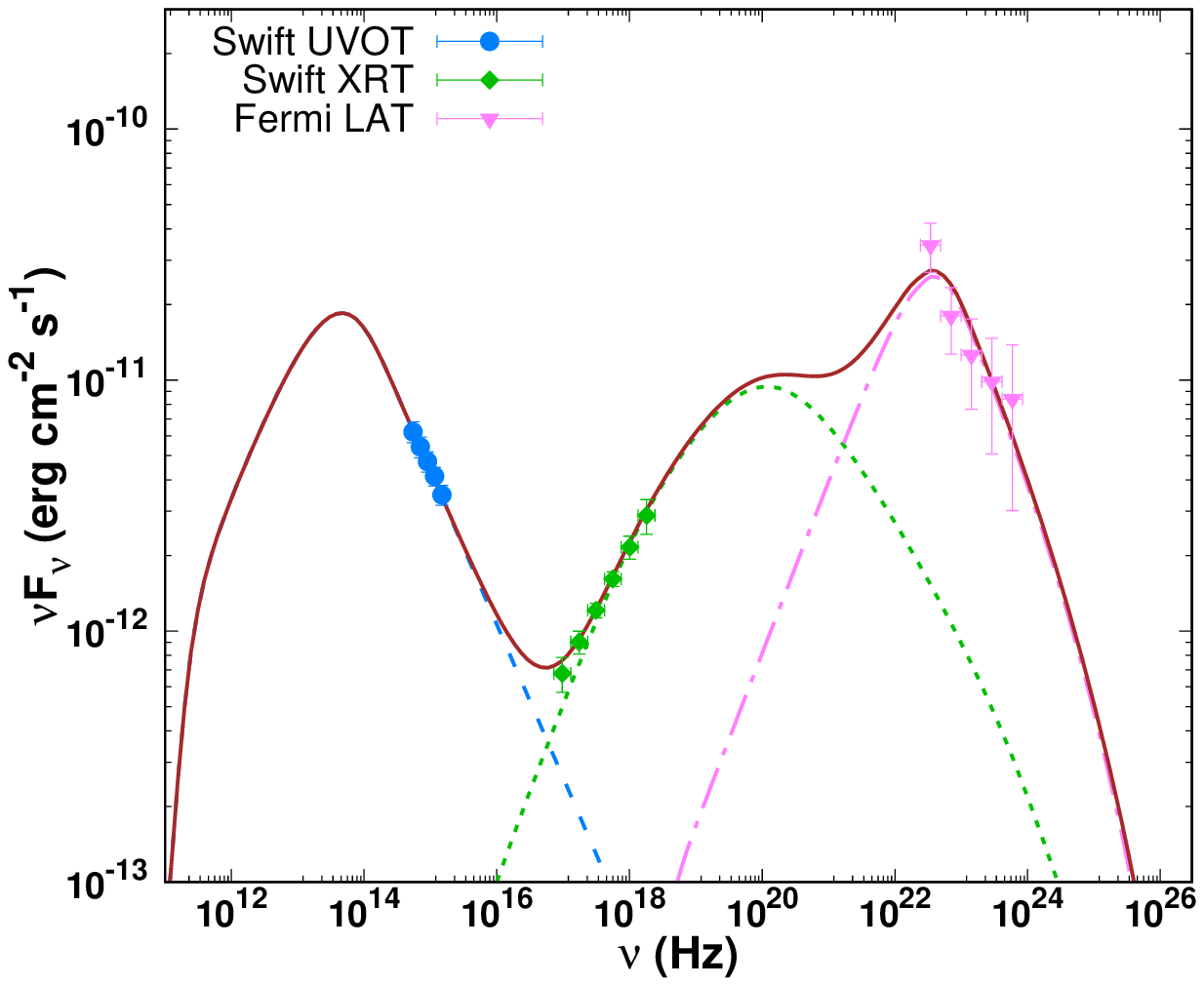}
		\caption{}
		\label{fig:e}
	\end{subfigure}
	\hspace{-2cm}
	\begin{subfigure}{0.49\textwidth}
		\includegraphics[width=0.55\textwidth, angle=-90]{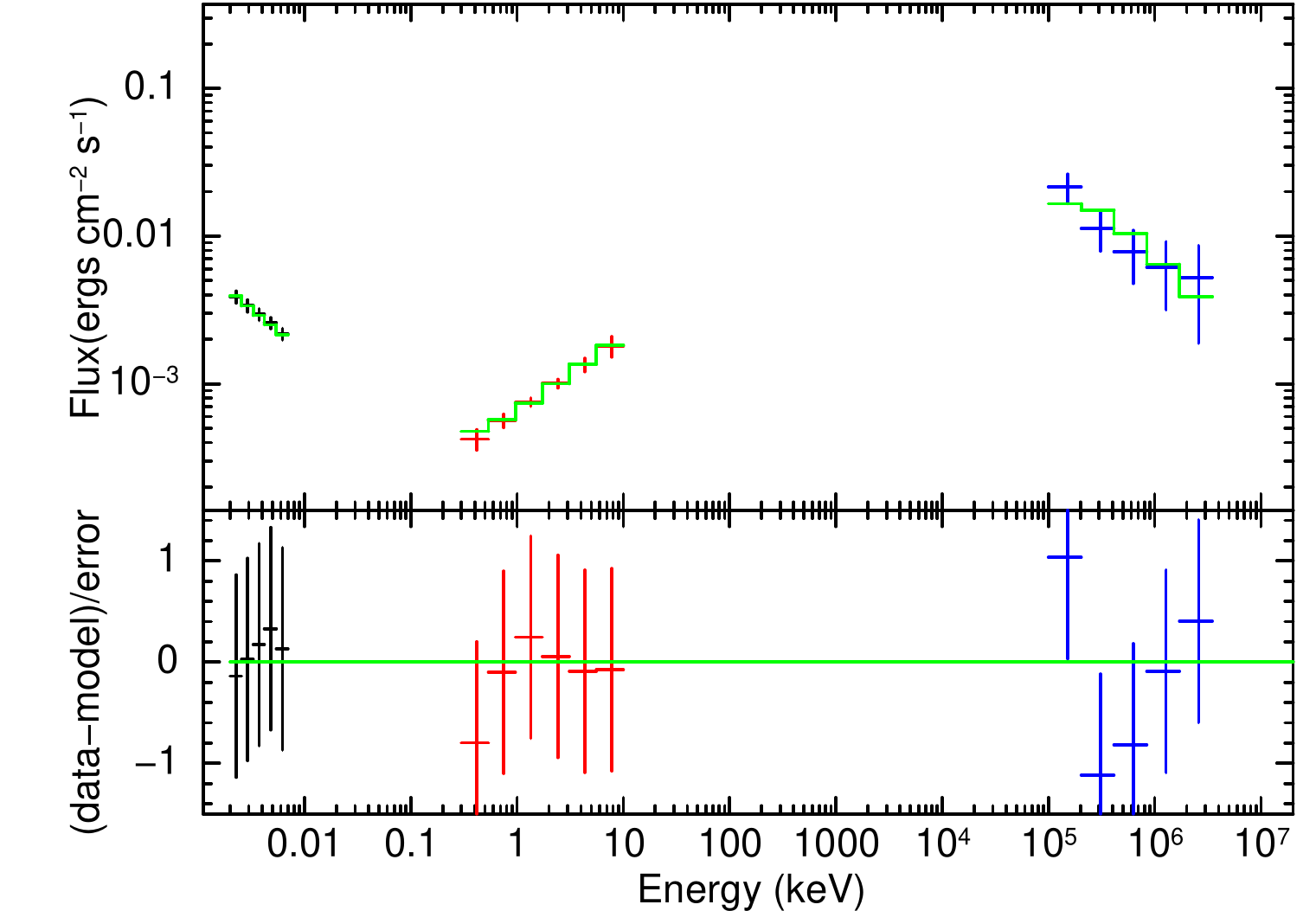}
		\caption{}
		\label{fig:f}
	\end{subfigure}
	\caption{Broadband SEDs along with best-fit models (left: a) flare 1, c) flare 2, and e) quiescent state. Right panels show the corresponding \emph{Xspec} plots showing the fit. In the left panels, dashed lines represents the synchrotron, dotted line represents the SSC and dotted dashed line the EC processes. The solid curve shows the total best-fit model.}
	\label{fig:sed2}
\end{figure*}

The spectral fit to the different activity states using these emission processes was first performed
by setting all parameters free except the ones mentioned above. However, the confidence intervals 
were obtained only for the parameters p,q, $\gamma_b$, B, and $U_{\rm target}$, since the iterations were unable to 
converge due to plausible degeneracies. In Table \ref{tab:sed}, we provide the details of the best-fit parameters, and the spectrum corresponding to these best-fit parameters 
are shown in Figure \ref{fig:sed2} with the right panels showing the output from \emph{XSpec}.
\begin{table*}
	\caption{Best fit values of the source parameters from broadband SED fitting.}
	\label{tab:sed}
	\setlength{\tabcolsep}{12pt}
	\begin{tabular}{lllll}
		\hline\hline
		Name of parameter& Symbol & Flare1 & Flare2 & Quiescent \\
		\hline\hline
		Low energy Particle index    &  $p $     &    2.3$\pm$0.04 &    1.8$\pm$0.05  & 1.6$\pm$0.1 \\ 
		High energy Particle index    &     $q$    &    4.4$\pm$0.1  &    4.8$\pm$0.2 & 4.2$\pm$0.1 \\ 
		Break Lorentz factor    &     $\gamma_b$  &   4542$\pm$508 &   2617$\pm$257 & 1230$\pm$ 69 \\ 
		Magnetic Field (G)  & $B$ & 0.86$\pm$0.03 & 0.80$\pm$0.01 & 0.68$\pm$0.02 \\ 
		Bulk Lorentz factor& $\Gamma$ & 28 & 16 & 13 \\ 
		Doppler Factor  &     $\delta$   &   28.6 &   24.4 & 21.5  \\ 
		Equipartition & $\eta$ & 3.0 & 3.0 & 3.0 \\
		External Compton Process & &EC-IR & EC-IR & EC-IR \\
		External photon energy density (erg/cm$^3$)& $U_{\rm target}$ & $(3.6\pm 0.01)\times10^{-4}$ & $(2.2\pm 0.01)\times10^{-4}$ & $(6.1 \pm 0.006 )\times10^{-5}$ \\
		\hline\hline
	\end{tabular}
\end{table*}

The observed fluxes at the three selected epochs were reasonably reproduced by the model using synchrotron, SSC and EC-IR emission processes. Comparing the SEDs in the Figure \ref{fig:sed1}, one can note that the flux has substantially increased across all the wavebands during flares. Our SED fit results suggest that this transition is associated with significant variations in most of the source parameters, including low and high energy particle indices, bulk Lorentz factor, $\gamma_b$, magnetic field $B$, and the target photon energy density $U_{\rm target}$. 
The variation in $U_{\rm target}$ across the activity states can be associated with the location
of the emission region from the central black-hole \cite{Ghisellini_2009}. The target photon energy density is higher for the flaring states compared to quiescent period, and this implies that during the flares, the emission region is more closer to central black-hole than in the case of quiescent state.
Similarly, the energy density for Flare 1 is larger than Flare 2, suggesting the location of the emission region may be closer to the central black hole for Flare 1. 
The SEDs also exhibit a significant difference in the 
spectral shape. For example, the X-ray spectrum during flare 1 has the contribution of synchrotron and SSC processes and shows a negative curvature, while for other
epochs the X-ray emission is predominantly due to SSC process.

Our broadband analysis suggests that the $\gamma$-ray emission from 4C\,31.03 can be due to the inverse Compton scattering of IR photons from the torus. This is consistent with our findings from the $\gamma$-ray observational results mentioned in previous sections. Further, the observed cooling time scale of the underlying electron distribution
for the EC emission can be calculated as \cite{2010MNRAS.405L..94T,2013ApJ...766L..11S},
\begin{align}
\label{eq:coolt}
t^{\rm obs}_c\,=\,\frac{3m_ec}{4\sigma_T U^{\prime}}\,\left( \frac{\nu_{\rm seed}^{\rm obs}\,\Gamma (1+z)}{\nu_{\rm IC}^{\rm obs}\, \delta}\right)^{\frac{1}{2}}
\end{align}
where, $U^{\prime}$ ($U_{\rm target}\,\,\Gamma^{2}$) is the target photon field energy density in the co-moving frame, 
$\nu_{\rm seed}^{\rm obs}$ ($2\times 10^{13}$Hz) is the observed seed photon frequency (for torus IR as the target photon field) and $\nu_{\rm IC}^{\rm obs}$ is the observed EC peak frequency.
The peak EC frequency is obtained from the best-fit values of the log-parabola spectral fit to the $\gamma$-ray 
observations \cite{2006A&A...448..861M}. In the case of flare 1, $\nu_{\rm IC}^{\rm obs}$ is equal to $1.2\times10^{23}$ Hz and using the best-fit values provided in Table \ref{tab:sed}, we 
obtained $t^{\rm obs}_c$ to be 7.3 hours. This is in close agreement with the decay time scales obtained from the temporal analysis of the source (\S\ref{sec:tvar}). 
\subsection{$\gamma$-ray flux distribution}
\begin{figure*}
	\centering
	\begin{subfigure}{0.49\textwidth}
		\includegraphics[width=0.75\textwidth]{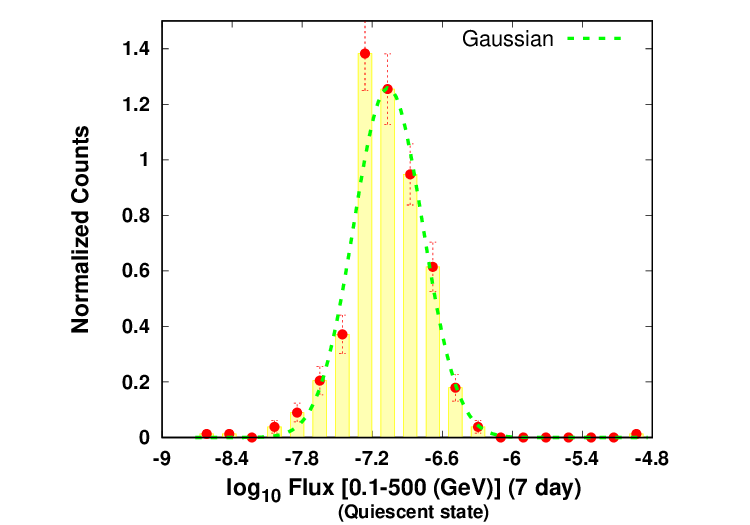}
		\caption{}
		\label{fig:a}
	\end{subfigure}
	\hspace{-2cm}
	\begin{subfigure}{0.49\textwidth}
		\includegraphics[width=0.75\textwidth]{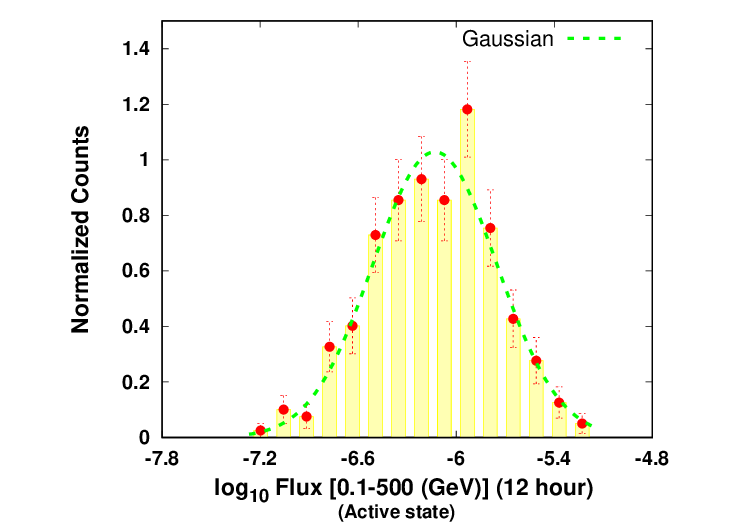}
		\caption{}
		\label{fig:b}
	\end{subfigure}
	
	\begin{subfigure}{0.49\textwidth}
		\includegraphics[width=0.75\textwidth]{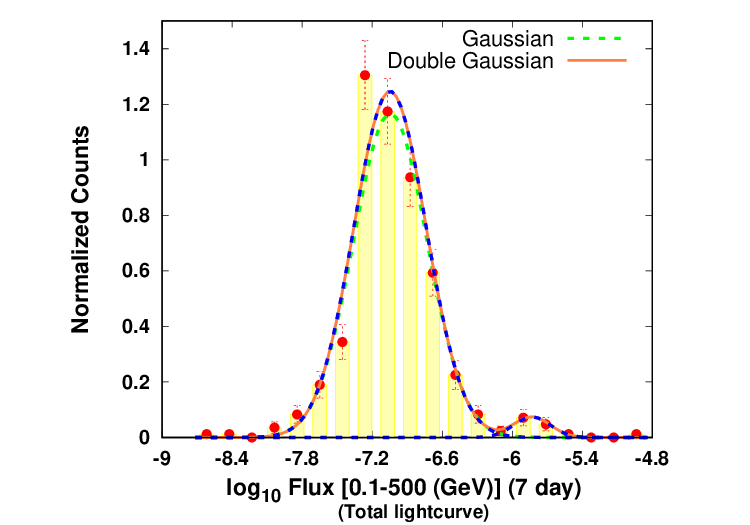}
		\caption{}
		\label{fig:c}
	\end{subfigure}
	\hspace{-2cm}
	\begin{subfigure}{0.49\textwidth}
		\includegraphics[width=0.75\textwidth]{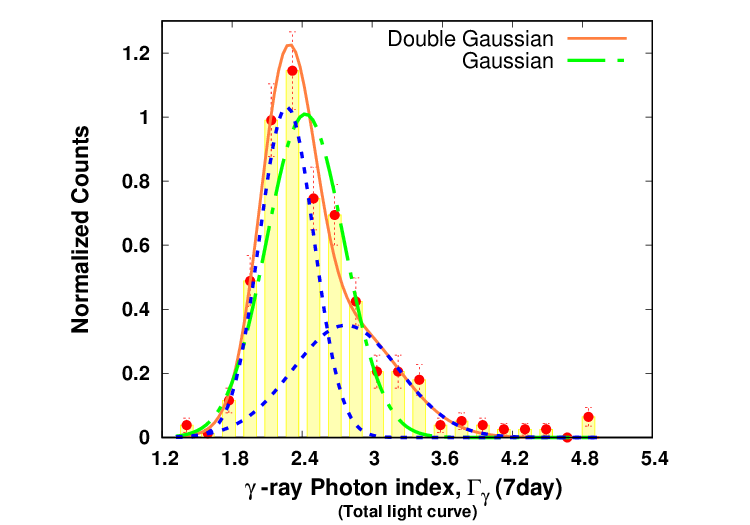}
		\caption{}
		\label{fig:d}
	\end{subfigure}
	
	\caption{Plots showing the histogram fitting of logarithm flux distributions obtained for a) Quiescent state, b) Active state,  c) Total lightcurve, and d) The index distribution obtained from the total 7 day binned lightcurve.}
	\label{hist1}	
\end{figure*}
	
The long-term flux variability of blazars are log-normal in nature and was reported by several authors \cite{2018Galax...6..135R,2003MNRAS.345.1271V,2019Galax...7...28R}. 
The log-normal variability is generally attributed when the underlying mechanism causing variability is multiplicative in nature,
with the perturbations propagating from accretion disc as an example (\cite{2005MNRAS.359..345U, 2012MNRAS.420..604N}).
However, the observed rapid variability in blazars supports a compact 
emission region which in turn demands the origin of variability to be associated with the processes happening in the jet \cite{Abdo_2010, 2015ApJ...808L..48P, 2016ApJ...824L..20A,2018ApJ...854L..26S}. When the narrow-band photon spectrum from the jet is defined by a power-law, a log-normal variability can also be an outcome of normal variation of the spectral index \cite{2018MNRAS.480L.116S,2020MNRAS.491.1934K}. In diffusive shock acceleration, the random fluctuations in the acceleration time scale induce normal variations in index which can result a a log-normal flux distribution \cite{2019Galax...7...28R}.  
Further, the blazars with steeper $\gamma$-ray spectrum have been found to be more variable since a steeper spectrum indicates higher relative amplitude fluctuations within shorter time intervals \cite{2020ApJ...891..120B}. Besides these, certain blazars also exhibit a double log-normal feature which may indicate that they are governed by two definite flux states \cite{2020MNRAS.491.1934K,2018RAA....18..141S,2023MNRAS.526.6364T}.

Being a bright $\gamma$-ray source after the prolonged low activity phase, 4C\,31.03 forms an ideal blazar to study the log-normal
flux and spectral variations. We obtained the histograms for the logarithm of fluxes and indices from the 7 day binned long-term 
lightcurve. To identify whether the source depicts two flux states, we first 
performed the histogram study using only the prolonged quiescent phase and then 
repeated with the inclusion of the recent active period. The histogram analysis reveals that the long-term quiescent phase flux alone can reasonably be reproduced by a single log-normal function. Conversely, a double log-normal function gives better representation for the the total flux distribution after including the active phase. However, in the double log-normal distribution, the Gaussian component corresponding to the high flux state is significantly suppressed due to the prolonged quiescent activity. 
Hence, we further explored the flux distribution of the active state using histogram obtained from 12 hour binned lightcurve (for better statistics). The active state flux distribution is well addressed by a log-normal distribution in the histogram study.
The histogram study,therefore supports that the blazar 4C\,31.03 is characterized by two definite flux states corresponding to its quiescent and active phases. 
The results of this study are given in Table \ref{tab:dist} and the histograms with the best-fit models are shown in Figure \ref{hist1}.
\begin{table*}
	\caption{Best fit parameter values of the PDF functions fitted to the logarithm of flux and index histogram. (PDF: GI stands for Gaussian and GII for double Gaussian)}
	\label{tab:dist}
	\setlength\tabcolsep{7pt}
	\begin{tabular}{l l l l l c c c  r }
		
		\hline
		\hline
		& Histogram  & PDF & $\mu_1$ & $\sigma_1$  & $\mu_2$ & $\sigma_2$  &     \multicolumn{1}{c}{$a_1$}  &    $\rm \chi^2/dof$ \\
		\hline
		7 day binned  & log10(Flux) & GI &-7.1$\pm$0.02  & 0.30$\pm$0.01  &  &   & 0.93$\pm$0.07  &  2.2\\
		(Quiescent state)& Index  & GI & 2.4$\pm$0.03 & 0.3$\pm$0.02 & &  & 0.84$\pm$ 0.08 &    3.9\\
		
		&  & GII& 2.3 $\pm$   0.03  & 0.23$\pm$0.04  & 2.8$\pm$0.2 & 0.45$\pm$0.08 & 0.60$\pm$0.2 &  1.8 \\
		\hline
		12 hour binned  & log10(Flux) & GI & -6.1$\pm$0.01 & 0.40$\pm$0.01  &  &  & 1.0$\pm$0.04 &  0.7 \\
		(Active)& Index  & GI &2.1$\pm$0.03 & 0.24$\pm$0.02 &  &  & 0.83$\pm$0.1 &  4.5\\
		&   & GII & 2.0$\pm$0.01 & 0.20$\pm$0.01 & 2.5$\pm$0.1  & 0.40$\pm$0.05 & 0.64$\pm$0.09 &  0.7\\
		\hline
		7 day binned  & log10(Flux) & GI & -7.0$\pm$0.02 & 0.31$\pm$0.02  &  &  & 0.92$\pm$0.07& 2.8\\
		(Total)& & GII & -7.0$\pm$0.02 & 0.31$\pm$0.01 & -5.8$\pm$ 0.07 & 0.16$\pm$ 0.06 & 1.0$\pm$0.01 &  2.2\\
		& Index  & GI &2.4$\pm$0.04 & 0.37$\pm$0.03 &  &  & 0.9$\pm$0.09 &  3.9\\
		
		& & GII & 2.3$\pm$0.04 & 0.21$\pm$0.04 & 2.6$\pm$0.08 & 0.50$\pm$0.04 & 0.45$\pm$0.1 &  1.8\\
		\hline
		\hline
	\end{tabular}
\end{table*}

We extended our study to understand the index variations by constructing their histograms (Figure \ref{hist1}(d)). 
Interestingly, the index distributions demanded a double Gaussian shape irrespective of the flux states.
A possible reason for this could be that the observed $\gamma$-ray 
spectrum falls around the peak of the Compton component and hence the spectrum is not a simple power-law.
During the different flux states the $\gamma$-ray peak shifts significantly and the integrated flux can be either governed by low-energy or the high-energy index.

The observed log-normal flux variability can be attributed the normal variation in the spectral index.
However, we find the index distributions during the flaring and quiescent states are better
represented by a double Gaussian function while the flux distributions during these states
suggested a single log-normal variability. Alternatively, the observed log-normality in blazars 
can also be attributed to the Fermi acceleration processes in the jet \cite{2019Galax...7...28R,2019ApJ...884...92X,2017A&A...598A..39H}. For instance, the quiescent period of the blazar may be dominated by one type of Fermi acceleration (turbulent/shock)
and at the flaring period, both could be at play \cite{2021MNRAS.505.2712D,2019ApJ...884..116L}. Consistently, the flux distribution of the total period reflects a double log-normal behavior.

To scrutinize our inference drawn from the histograms, we tested the hypothesis of normality for the distributions of $\gamma$-ray fluxes and indices using Anderson-Darling (AD) statistics. 
The test results are provided in Table \ref{tab:ad}. According to the AD test performed, the logarithmic flux distribution of the total lightcurve deviates significantly from a single Gaussian behavior. This result is consistent with conclusions drawn from the histogram study which supports a double log-normal variability.
 The AD test performed on the active phase flux alone supports a log-normal variability 	
in agreement with the corresponding histogram analysis. 
Interestingly, the AD test for the quiescent phase
suggests slight deviation from the log-normal behavior which is contrary to the conclusion
drawn from the histogram study. A plausible reason could be, the quiescent period may 
also contain minor flares associated with different physical processes. The presence of such flares 
deviate the variability behavior from log-normal in case of AD test. However, in case of histogram 
study, these minor flares may be over shadowed by the prolonged quiescent phase and hence suggest a log-normal variability.
\begin{table*}
	\caption{Results of the Anderson Darling test done for the flux/index distributions. The corresponding distributions will follow a normal only when the statistics becomes lesser than critical value.}
	\label{tab:ad}
	\begin{tabular}{c c c c c}
		\hline
		\hline
		&  \multicolumn{1}{c}{Number of }   &  \multicolumn{1}{c}{Normal (Flux)}            &     \multicolumn{1}{c}{Normal (log Flux)} & \multicolumn{1}{c}{Normal (Spectral index)}\\ 
		& data points & AD(critical value) &  AD(critical value) &  AD(critical value) \\ \hline  
		
		7 day binned Quiescent state   & 1354  & 124 (0.779) & 2.44 (0.779) & 12.8 (0.779) \\
		12 hour binned active state & 326 & 15.2 (0.776) & 0.48 (0.776) &  10.9 (0.776)\\
		7 day binned total & 5850 & 118 (0.78) & 5.9 (0.78) & 13.3 (0.78)\\							   
		\hline	\hline
	\end{tabular}
\end{table*}
\section{Summary}
The blazar 4C\,31.03 has recently been reported by \emph{Fermi}-LAT for exhibiting an exceptionally high $\gamma$-ray activity for the first time. In this paper, we performed a detailed study of this major outburst through temporal and spectral analysis. 
We used publicly available observations by \emph{Fermi}-LAT, \emph{Swift}-XRT, and UVOT instruments for this study.

We performed the temporal study after obtaining $\gamma$-ray lightcurves with 7 day, 1 day, and 12 hour binning. 
Through the statistical analysis of lightcurves using Bayesian blocks, we identified 3 epochs featuring prominent flare. 
The shortest variability timescale of $5.5\pm 0.7$ hours has been found from the variability analysis performed using 12 hour binned lightcurve. The comparison of fluxes in the energy ranges 0.1-1 GeV and 1-500 GeV indicates that this outburst can be majorly 
caused by the low-energy electrons in the particle distribution. The $\gamma$-ray SEDs corresponding to 
flaring and quiescent states have been studied using various models, revealing that the spectra during the flares exhibit 
significant curvature/break.

The maximum photon energy, detected positionally consistent with the source, is $\sim$ 82 GeV, and this implies that the emission region may be situated outside the BLR. Using the pair production opacity argument and considering variability time scale, we obtained the value of minimum jet Doppler factor corresponding to the flaring state as 17 and 13 for the quiescent state. The upper limit on the emission region size has been estimated as $\sim\,10^{16}$ cm. 

Further, we performed a detailed statistical broadband SED study of 3 multi-wavelength spectra representing flaring and quiescent activities. Using a one-zone leptonic model that incorporates synchrotron, SSC, and EC emission processes, the observed fluxes at considered energy bands have been successfully reproduced. Interestingly, the observed $\gamma$-ray fluxes are well accounted for by attributing them with the external Compton process involving IR photons from the torus. Further the value of observed cooling timescale, $t^{\rm obs}_c$ estimated from the broadband SED results, closely aligns with the values of time scales obtained from the variability analysis. Our study supports the idea that the $\gamma$-ray production in 4C\,31.03 happens beyond the BLR, but within the molecular torus.
 
The study of the variability of the $\gamma$-ray emission from 4C\,31.03 has been extended by performing a comprehensive long-term distribution study. The $\gamma$-ray emission from 4C\,31.03 exhibits log-normal variability and showcase a double flux state corresponding to the prolonged quiescent and the recent active phases. The observed log-normal flux distribution can be attributed to the Fermi acceleration processes inside the jet. 
The index distribution also deviates significantly from normal and features a double normal variability. This may infer that the $\gamma$-ray spectrum can be more accurately described by two photon indices rather than a single one.

\section*{Acknowledgements}
The authors thank Zahir Shah and Vaidehi S. Paliya for their valuable suggestions.	
This work has made use of observations from NASA's Fermi gamma-ray and Swift telescopes. The data used in this work were obtained from archives available at Fermi gamma-ray telescope support centre, a service of the Goddard Space Flight Center and the Smithsonian Astrophysical Observatory and NASA's High Energy Astrophysics 
Science Archive Research Center (HEASARC).
AT is thankful to UGC-SAP and FIST 2 (SR/FIST/PS1-159/2010) (DST, Government of India) 
for the research facilities provided in the Department of Physics, University of Calicut.

\section*{Data Availability}
The data used in this paper are publicly available from the
archives at \url{https://heasarc.gsfc.nasa.gov/} and \url{https://Fermi.gsfc.nasa.gov/}.




\bibliographystyle{elsarticle-num-names} 
\bibliography{cas-refs}



\end{document}